\pdfoutput=1
\documentclass[sigconf,pbalance,nonacm]{acmart}

\usepackage{amsmath,amsfonts}
\usepackage{textcomp}
\usepackage{lipsum}
\usepackage[dvipsnames,table]{xcolor}
\usepackage{colortbl}
\usepackage{subcaption,adjustbox}
\usepackage{endnotes,microtype,xspace,graphicx,fancyvrb,multirow}
\usepackage{booktabs}
\usepackage{longtable}
\usepackage{array,underscore,relsize}
\usepackage[T1]{fontenc}

\usepackage{enumitem}
\setlist[itemize]{topsep=1pt,itemsep=0pt,parsep=0pt}
\setlist[enumerate]{topsep=1pt,itemsep=0pt,parsep=0pt}
\usepackage[labelfont=bf,skip=5pt]{caption}
\usepackage{tabularx}
\usepackage{fp}
\usepackage{siunitx}
\usepackage{comment}

\usepackage{listings}
\usepackage{float}
\floatstyle{plain}
\newfloat{listing}{htbp}{lol}
\lstdefinestyle{codefootnotesize}{%
  basicstyle=\footnotesize\ttfamily,
  breaklines=true,
  breakatwhitespace=false,
  columns=flexible,
  keepspaces=true,
  lineskip=-1pt,
}
\lstdefinestyle{codesmall}{%
  basicstyle=\small\ttfamily,
  breaklines=true,
  breakatwhitespace=false,
  columns=flexible,
  keepspaces=true,
}
\lstdefinestyle{pbfuzz-code}{
  basicstyle=\footnotesize\ttfamily,
  columns=fullflexible,
  keepspaces=true,
  showstringspaces=false,
  breaklines=true,
  breakindent=0pt,
  breakautoindent=false,
  lineskip=-1.25pt,
  upquote=true
}
\lstdefinestyle{pbfuzz-c}{
  style=pbfuzz-code,
  language=C,
  numbers=left,
  numberstyle=\tiny\color{gray},
  numbersep=6pt,
  xleftmargin=2em,
  keywordstyle=\bfseries\color[rgb]{0.00,0.50,0.00},
  commentstyle=\itshape\color[rgb]{0.24,0.48,0.48},
  stringstyle=\color[rgb]{0.73,0.13,0.13},
  deletekeywords={void,char,int},
  morekeywords=[2]{void,char,int},
  keywordstyle=[2]\color[rgb]{0.69,0.00,0.25},
  emph={xmlSnprintfElementContent},
  emphstyle=\color[rgb]{0.00,0.00,1.00}
}
\lstdefinelanguage{pbfuzz-json}{
  sensitive=true,
  alsoletter={"0123456789},
  morekeywords={
    "complexity","content_model_type","description","element_name_length",
    "element_prefix_length","id","input_constraints","max","min",
    "nesting_depth","num_elements","precondition_ids","statement","status",
    "type","values"
  },
  keywordstyle=\bfseries\color[rgb]{0.00,0.50,0.00},
  identifierstyle=\color[rgb]{0.73,0.13,0.13}
}
\lstdefinestyle{pbfuzz-json}{
  style=pbfuzz-code,
  language=pbfuzz-json
}

\usepackage{algorithm}
\usepackage{algpseudocode}
\usepackage{multicol}

\usepackage{tikz}
\usetikzlibrary{arrows.meta,patterns,calc,positioning}

\def\Snospace~{\S{}}

\newcommand{\cc}[1]{\mbox{\smaller[0.8]\texttt{#1}}}

\def\BibTeX{{\rm B\kern-.05em{\sc i\kern-.025em b}\kern-.08em
    T\kern-.1667em\lower.7ex\hbox{E}\kern-.125emX}}

\definecolor{mGreen}{rgb}{0,0.6,0}
\definecolor{mGray}{rgb}{0.5,0.5,0.5}
\definecolor{mPurple}{rgb}{0.58,0,0.82}
\definecolor{backgroundColour}{rgb}{0.95,0.95,0.92}

\newcommand{\sys}{\mbox{\textsc{PBFuzz}}\xspace}
\newcommand{\weburl}[2]{\href{#1}{\textit{\textcolor{ACMDarkBlue}{#2}}}}

\newcommand{\cursor}{\mbox{\textsc{cursor}}\xspace}
\newcommand{\cursorFB}{\mbox{cursor$^{\dagger}$}\xspace}
\newcommand{\cursorTools}{\mbox{cursor$^{\ddagger}$}\xspace}

\usepackage{amsthm}
\theoremstyle{definition}

\newcommand{\PP}[1]{
  \par
  \textit{#1.}
}

\usepackage[most]{tcolorbox}
\newtcolorbox{findingbox}{
  colback=gray!10,
  colframe=black!70,
  arc=3pt,
  boxrule=0.5pt,
  left=6pt, right=6pt, top=4pt, bottom=4pt,
}

\makeatletter
\AtBeginDocument{%
  \let\orig@ref\ref
  \renewcommand{\ref}[1]{%
    \ifnum\pdfstrcmp{#1}{TotPages}=0 15\else\orig@ref{#1}\fi%
  }%
}
\makeatother

\copyrightyear{2026}
\acmYear{2026}
\setcopyright{cc}
\setcctype{by-nc-nd}
\acmConference[CCS '26]{Proceedings of the 2026 ACM SIGSAC Conference on Computer and Communications Security}{November 15--19, 2026}{The Hague, Netherlands}
\acmBooktitle{Proceedings of the 2026 ACM SIGSAC Conference on Computer and Communications Security (CCS '26), November 15--19, 2026, The Hague, Netherlands}
\acmDOI{10.1145/3830454.3832663}
\acmISBN{979-8-4007-2871-6/2026/11}

\begin{document}

\title{PBFuzz: Agentic Directed Fuzzing for PoV Generation}

\author{Haochen Zeng}
\email{hzeng013@ucr.edu}
\affiliation{
  \institution{University of California, Riverside}
  \city{Riverside}
  \state{California}
  \country{USA}
}

\author{Andrew Bao}
\email{bao00065@umn.edu}
\affiliation{
  \institution{University of Minnesota, Twin Cities}
  \city{Minneapolis}
  \state{Minnesota}
  \country{USA}
}

\author{Jiajun Cheng}
\email{jchen1192@ucr.edu}
\affiliation{
  \institution{University of California, Riverside}
  \city{Riverside}
  \state{California}
  \country{USA}
}

\author{Chengyu Song}
\email{csong@cs.ucr.edu}
\affiliation{
  \institution{University of California, Riverside}
  \city{Riverside}
  \state{California}
  \country{USA}
}

\begin{abstract}
Proof-of-Vulnerability (PoV) input generation is a critical task in software security
and supports downstream applications such as bug reproduction and patch validation.
Generating a PoV input requires solving two sets of constraints:
(1) \emph{reachability constraints} for reaching vulnerable code locations,
and (2) \emph{triggering constraints} for activating the target vulnerability.
Existing approaches, including directed greybox fuzzing and LLM-assisted fuzzing,
struggle to efficiently satisfy these constraints.

This work presents an agentic method that mimics human experts.
Human analysts iteratively study code to extract semantic reachability and triggering constraints,
form hypotheses about PoV triggering strategies, encode them as test inputs,
and refine their understanding using debugging feedback.
We automate this process with an agentic directed fuzzing framework called \sys.
\sys tackles four challenges in agentic PoV generation:
autonomous code reasoning for semantic constraint extraction,
custom program-analysis tools for targeted inference,
persistent memory to avoid hypothesis drift,
and property-based testing for efficient constraint solving while preserving input structure.
Experiments on the Magma benchmark show strong results.
\sys triggered 57 vulnerabilities, surpassing all baselines,
and uniquely triggered 17 vulnerabilities not exposed by existing fuzzers.
\sys achieved this within a 30-minute budget per target,
whereas conventional approaches were allotted 24 hours.
Median time-to-exposure was 339 seconds for \sys versus 8,680 seconds for AFL++ with CmpLog,
giving a $25.6\times$ efficiency improvement with an API cost of \$1.83 per vulnerability.
In real-world applications, \sys reproduced three \cc{FFmpeg} 1-day CVEs that had no public PoVs
and identified three previously unknown Bento4 vulnerabilities.
\end{abstract}
\thanks{Accepted to ACM CCS 2026.}

\ccsdesc[500]{Software and its engineering~Software testing and debugging}
\ccsdesc[300]{Theory of computation~Program analysis}
\ccsdesc[100]{Security and privacy~Software security engineering}
\ccsdesc[100]{Security and privacy~Vulnerability scanners}

\keywords{Directed Fuzzing; Property-Based Testing; AI for Security}

\maketitle

\section{Introduction}
\label{sec:intro}

Generating a proof-of-vulnerability (PoV) input that triggers a vulnerability
is a critical task in software security.
The state-of-the-art (SOTA) approach for this task is directed greybox fuzzing,
where, instead of trying to maximize overall code coverage
\cite{fioraldi2020aflpp,honggfuzz,serebryany2016continuous},
directed greybox fuzzers (DGFs) aim to generate test inputs to exercise \emph{specific functionalities} or
\emph{vulnerable code locations} in the program under test (PUT)~\cite{weissberg2024sok}.
DGFs have been applied in various scenarios, including
static analysis alarms verification~\cite{bao2025alarms},
bug reproduction~\cite{kim2023dafl, lee2021constraint, huang2022beacon},
patch validation~\cite{bohme2017directed, chen2018hawkeye, du2022windranger},
commit testing~\cite{xiang2024critical},
and regression testing~\cite{zhu2021regression}.

DGFs typically follow a two-step process.
First, they perform static analysis to extract guidance (e.g., a distance metric)
on how to reach the target location(s)
from the PUT's low-level program structures such as the control-flow graph (CFG),
definition-use graph (DUG), and data-flow graph~\cite{weissberg2024sok}.
During fuzzing, DGFs use this guidance to evaluate generated inputs,
prioritize scheduling of seeds that are more likely to reach the target,
and perform more mutations on those seeds.
Intuitively, one might expect DGFs to excel at reaching target locations faster.
However, recent research~\cite{bao2025alarms,she2024fox,huang2023titan,geretto2025libaflgo} has shown that
DGFs struggle to generate PoV inputs; in some cases, they even underperform
coverage-guided fuzzers like AFL++~\cite{fioraldi2020aflpp, fioraldi2022libafl}.
In fact, our experimental results (\autoref{fig:fuzzer-comparison}) on the Magma benchmark show that
AFLGo triggered only 41 CVEs, while AFL++ with Cmplog triggered 49.
These results suggest that, despite their goal-oriented design,
DGFs are neither particularly effective nor efficient at generating PoV inputs.

With recent advancements in large language models (LLMs),
there has been growing interest in leveraging LLMs' ability to
generate syntactically valid and semantically plausible inputs
to address diverse challenges in fuzzing, including:
format-aware input generation~\cite{meng2024large,zhang2025low,tu2025large},
coverage plateaus~\cite{shi2024harnessing,xia2024fuzz4all},
semantics-aware mutations~\cite{deng2023large,xia2024fuzz4all},
and vulnerability-targeted exploration~\cite{xu2025directed,feng2025fuzzing,zhu2025locus}.
A natural question therefore arises: can LLMs help PoV input generation?
Our preliminary experiments with one-shot chatbot-style prompting showed mixed results:
LLMs generated PoV inputs for simple vulnerabilities
(13 out of 129 CVEs from Magma~\cite{hazimeh2020magma}),
but struggled with complex ones (details in \autoref{sec:eval:ablation}),
motivating more capable approaches.

Fundamentally, a PoV input must satisfy two sets of constraints:
(1) \emph{reachability constraints} enabling execution to reach target locations,
and (2) \emph{triggering constraints} activating the vulnerability.
Successfully generating PoV inputs requires solving three key challenges:
{\bf C1}. accurately extracting these constraints,
{\bf C2}. efficiently solving them,
and {\bf C3}. effectively encoding solutions into test inputs.
Our investigation revealed three main limitations in applying LLMs to solve these challenges:
{\bf L1}. \emph{long-context reasoning} challenges when analyzing large codebases
and long code paths;
{\bf L2}. \emph{hypothesis drift} where iterative feedback fails to escape incorrect constraint sets;
and {\bf L3}. \emph{slow, imprecise constraint solving} through direct LLM inference.

\PP{Our Approach}
We propose \sys, an \textit{agentic approach} to unlock LLMs' reasoning, planning
and coding capabilities for effective and efficient PoV input generation.
Rather than treating LLMs as auxiliary components to traditional fuzzers (and static analysis),
\sys leverages an LLM as an autonomous agent that can perform high-level code reasoning,
planning, tool orchestration, code generation, and self-reflection to \emph{iteratively}
\begin{enumerate}[label=(\arabic*),leftmargin=*,topsep=0pt]
\item analyze code and extract reachability and triggering constraints,
\item design solving strategies and encode them as parameterized input generators,
\item leverage property-based testing~\cite{claessen2000quickcheck,goldstein2024property}
to systematically search the input space for PoV inputs, and
\item collect fine-grained execution feedback to validate and refine its hypotheses.
\end{enumerate}
This agentic design allows us to address the aforementioned limitations.
First, instead of overwhelming LLMs with extensive code snippets and
asking them to identify critical elements, LLM agents can explore the codebase
based on their own hypotheses regarding reachability and triggering conditions.
As highlighted in our case studies (\autoref{sec:eval:strengths-limitations}),
this approach significantly improves the accuracy of constraint extraction
and the efficiency of solving.
Second, LLM agents can seek custom fine-grained feedback to refine their
hypotheses; with persistent memory, they can also abandon
their current plans and backtrack to pursue new hypotheses.
Finally, with property-based testing~\cite{claessen2000quickcheck,goldstein2024property},
LLM agents write custom solvers to efficiently search the input space for PoVs.

Indeed, LLM agents have been explored in security domains~\cite{shahriar2025survey},
including DARPA's AIxCC~\cite{aicyberchallenge},
alongside coding agents like OpenAI Codex~\cite{chen2021codex},
Claude Code~\cite{anthropic2024claude}, and Cursor~\cite{cursor2024}.
However, we found off-the-shelf agents also struggle with PoV generation due to
four key operational challenges.
{\bf OC1}. \textit{Dynamic Hypothesis Validation:}
agents must develop hypotheses about input constraints,
then autonomously gather program information to validate them~\cite{yao2022react}.
{\bf OC2}. \textit{Persistent Memory Management:}
long-horizon tasks risk memory drift,
where agents lose track of validated reasoning~\cite{shinn2023reflexion}.
{\bf OC3}. \textit{Fine-Grained Execution Feedback:}
coarse-grained coverage feedback is insufficient for hypothesis refinement;
cascading reasoning errors in agentic systems require timely detection~\cite{zhu2025llm}.
{\bf OC4}. \textit{Efficient Constraint Solving:}
LLM inference is orders of magnitude slower than fuzzing~\cite{yang2025hybrid},
and LLMs lack precision in solving complex constraints~\cite{jiang2024towards}.
Our four-phase workflow design (\autoref{fig:arch}),
custom toolset, and memory management address these challenges.

To demonstrate the capability of our approach,
we implemented a prototype of \sys based on the cursor-cli tool
and open-sourced it at
\weburl{https://github.com/sgzeng/pbfuzz}{github.com/sgzeng/pbfuzz}.
Experiments on the Magma benchmark~\cite{hazimeh2020magma} show that
our agentic approach significantly outperformed previous solutions.
\sys successfully triggered 57 out of the 129 CVEs,
including 17 CVEs that none of the previous approaches triggered.
More critically, \sys achieved these results with a single trial and a 30-minute budget,
while baseline fuzzers were run with 10 trials of 24 hours each.
Our ablation study also confirms the contributions of each design component
and major innovations.
\sys also discovered three previously unknown vulnerabilities in Bento4,
a widely-deployed MP4 processing library.

\emph{Contributions.}
This paper makes three key contributions:
\begin{itemize}[leftmargin=*,topsep=0pt]

\item We present a novel approach for effective and efficient PoV input generation,
where we first conduct code analysis to extract semantic-level reachability
and triggering constraints, then leverage property-based testing to solve
the constraints at the input-space level.

\item We introduce an agentic framework that realizes and automates our approach.
Our agent, \sys, employs a custom four-phase workflow that features
autonomous code reasoning, on-demand tool orchestration,
persistent memory management, fine-grained execution feedback,
and property-based test generation.

\item We provide experimental validation of our approach on the Magma benchmark,
which shows that \sys outperformed all baselines, including traditional fuzzers,
LLM-assisted fuzzers, and off-the-shelf coding agents.

\end{itemize}

\section{Background}
\label{sec:background}

\subsection{Fuzzing and Directed Fuzzing}

Coverage-guided greybox fuzzing (CGF) is an automated software testing technique
that dynamically feeds mutated inputs to programs to discover vulnerabilities.
CGF employs lightweight instrumentation to collect runtime code coverage feedback,
guiding input generation toward unexplored program paths through evolutionary mutation.
Tools such as AFL~\cite{afl} and libFuzzer~\cite{serebryany2016continuous} have discovered thousands of vulnerabilities.

Despite its success, CGF blindly extends code coverage without prioritization.
Only a small fraction of source code contains vulnerabilities
(e.g., merely 3\% of Mozilla Firefox files~\cite{weissberg2024sok}).
Blindly maximizing coverage wastes resources on irrelevant code regions,
limiting efficiency for reaching specific vulnerability targets.

Directed greybox fuzzing (DGF) addresses this by focusing testing resources on specific target locations~\cite{bohme2017directed}.
Unlike CGF, which explores execution paths indiscriminately,
DGF prioritizes seeds closer to reaching predetermined targets
such as recently patched code or sensitive operations~\cite{weissberg2024sok}.
DGF computes distance metrics between current program locations and targets
to guide seed prioritization effectively~\cite{weissberg2024sok}.

\subsection{Property-Based Testing}
Software testing traditionally relies on manually crafted test cases that verify specific 
input-output pairs. While intuitive, this approach suffers from limited coverage and 
difficulty anticipating edge cases. Property-based testing (PBT) offers a different 
paradigm: instead of writing individual test cases, developers specify general properties 
that their programs should satisfy, and automated tools generate extensive diverse 
inputs to verify these properties~\cite{mayhem-pbt}.
Modern PBT frameworks have evolved significantly since QuickCheck's introduction in 
Haskell~\cite{claessen2000quickcheck}. 
Contemporary implementations include Hypothesis~\cite{maciver2019hypothesis} for Python, 
Google FuzzTest~\cite{fuzztest} for C++, and proptest~\cite{proptest} for Rust. 

PBT frameworks enable expressing triggering constraints
as typed parameter domains that the framework systematically explores~\cite{mayhem-useful-properties,goldstein2024property}.

\subsection{Property-Based Directed Fuzzing}
PBT and fuzzing are complementary approaches
(detailed comparison in \autoref{tab:pbt-vs-fuzzing}):
PBT operates on typed parameter domains,
while fuzzing mutates raw byte arrays.
Triggering vulnerabilities can be viewed as
violating safety properties where sanitizers act as implicit oracles.
This transforms directed fuzzing into specialized property-based testing
with two key components:
\textbf{(1) Preconditions} enabling execution to reach target locations,
and \textbf{(2) Triggering Conditions} activating the vulnerability.
Finding PoV inputs becomes discovering counterexamples
satisfying both reachability and triggering constraints.
\begin{table}[!htbp]
\centering
\caption{Property-Based Testing vs. Fuzzing}
\label{tab:pbt-vs-fuzzing}
\small
\begin{tabular}{l|p{86.21pt} p{86.21pt}}
\toprule
\textbf{Aspect} & \textbf{Property-Based Testing} & \textbf{Fuzzing} \\
\midrule
Goal & Verify invariant properties & Discover vulnerabilities \\
Bug Oracle & Explicit assertions & Implicit sanitizers \\
Input Space & Typed parameter domains & Raw byte arrays \\
Scope & Functional correctness & End-to-end testing \\
Duration & Seconds to minutes & Hours to days \\
\bottomrule
\end{tabular}
\end{table}

However, PBT faces a fundamental challenge:
synthesizing generators encoding complex structural preconditions~\cite{goldstein2024property}.
Real-world constraints demand deep domain knowledge---well-formed XML,
balanced red-black trees, valid S-expressions---
requiring substantial engineering effort~\cite{goldstein2024property}.
We propose an agentic approach addressing this bottleneck.
Our LLM agent \sys conducts program analysis,
inspects execution states, and refines hypotheses through feedback
to extract implicit constraints from entry points to vulnerability sites.
The agent synthesizes typed parameter domains and generator functions
encoding learned constraints as testable input domains.

\section{Motivation}
\label{sec:motivation}

\begin{figure}[bt]
  \begin{lstlisting}[style=pbfuzz-c]
void xmlSnprintfElementContent(char *buf, int size, xmlElementContentPtr content, int englob) {
  int len;
  if (content == NULL) return;
  len = strlen(buf);
  if (size - len < 50) {
    if ((size - len > 4) && (buf[len - 1] != '.'))
      strcat(buf, " ...");
    return;
  }
  if (englob) strcat(buf, "(");
  switch (content->type) {
    case XML_ELEMENT_CONTENT_ELEMENT:
      if (content->prefix != NULL) {
        strcat(buf, (char *) content->prefix);//BUG
        strcat(buf, ":");
      }
      if (content->name != NULL)
        strcat(buf, (char *) content->name);//BUG
      break;
    case XML_ELEMENT_CONTENT_SEQ:
      xmlSnprintfElementContent(buf,size,content->c1,1);
      strcat(buf, " , ");
      xmlSnprintfElementContent(buf,size,content->c2,0);
      break;
    case XML_ELEMENT_CONTENT_OR:
      xmlSnprintfElementContent(buf,size,content->c1,1);
      strcat(buf, " | ");
      xmlSnprintfElementContent(buf,size,content->c2,0);
      break;
  }
  if (englob) strcat(buf, ")");
}
\end{lstlisting}
\caption{CVE-2017-9047 buffer overflow in libxml2's \cc{xmlSnprintfElementContent()} function.
  The bounds check uses stale buffer length \cc{len} instead of updated \cc{strlen(buf)}
  after appending namespace prefixes. %
  }
  \label{fig:cve-2017-9047}
\vspace{-1.5em}
\end{figure}

\autoref{fig:cve-2017-9047} shows a buffer overflow in the libxml2 project.
The bug occurs when processing namespace prefixes: after concatenating the prefix (line~14),
the function appends the element name (line~18) without refreshing \cc{len},
so the remaining-space check at lines~5--6 still reflects the pre-prefix buffer state.
Deeply nested content models trigger recursive accumulation via lines~21--29.

Generating a proof-of-vulnerability (PoV) input for this bug requires satisfying four types of constraints.
The first three are \emph{reachability constraints} to reach the vulnerable code,
and the last one is the \emph{triggering constraint} to actually overflow the buffer.
(1) \emph{format-level}: the PoV input must be a syntactically valid XML with well-formed
Document Type Definition (DTD) ELEMENT declarations.
(2) \emph{structure-level}: the DTD content models must use nested SEQ/OR operators like \cc{(a,(b|(c,d)))}
rather than flat lists like \cc{(a,b,c,d)} to trigger recursive accumulation.
(3) \emph{namespace-level}: elements must have prefixes (\cc{prefix:name}) to reach lines~13--18.
(4) \emph{depth-level}: nesting depth must reach approximately 10 levels to accumulate sufficient buffer
consumption (70-character prefixes/names per level) to exhaust the 5000-byte buffer.

Existing approaches all struggle on this CVE.
AFL++ required approximately 20 hours, with a 60\% success rate.
SOTA LLMs (GPT-5, Claude 4.5) with source code context extracted with static
reachability analysis (from directed fuzzers)
produced inputs reaching the target but none triggered the vulnerability.
Off-the-shelf Cursor Agent with Claude Sonnet 4.5 successfully triaged
the vulnerability but failed to generate the PoV.

While our preliminary experiments were not fully successful,
they revealed a promising opportunity:
LLM agents possess semantic reasoning capabilities
that traditional approaches struggle with.
In other words, they can generate \emph{better semantic-level guidance} than
low-level distance metrics and path constraints.
But they lack the ability to efficiently solve the constraints in the input space.
Our key insight is that \emph{property-based testing (PBT) provides
this missing bridge between semantic analysis and constraint solving}.
Rather than generating inputs through expensive LLM inference,
our agent synthesizes \emph{parameterized input generators}
encoding preconditions as typed parameter spaces,
enabling high-throughput exploration with semantic guidance.

Applied to CVE-2017-9047, code exploration with runtime
debugging feedback %
guided our agent to %
design a five-dimensional input parameter space
including \cc{nesting\_depth}, \cc{element\_prefix\_length}, \cc{element\_name\_length},
\cc{content\_model\_type}, and \cc{num\_elements}.
Systematic sampling through PBT discovers the triggering configuration in 3.4 min:%
depth 10 with 70-character prefixes/names successfully triggers the vulnerability.

\section{Design of \sys}
\label{sec:design}

\subsection{Overview}
\label{sec:overview}

\subsubsection{Prerequisites}
\sys targets a narrower problem than end-to-end bug discovery.
Analogous to directed fuzzing, our focus is:
\emph{given a bug predicate and execution signals,
can an agent translate semantic constraints into a PoV input?}
This mirrors the standard directed-fuzzer setup and is
directly applicable to bug reproduction and patch validation.
We assume the following two prerequisites:
\begin{itemize}[leftmargin=*,topsep=0pt]
  \item \textbf{Bug predicate.}
  An existing analysis---manual inspection,
  patch diffing, or static checking---has identified
  the violated safety property at the target site.
  \item \textbf{Reach/trigger signals.}
  The PUT is instrumented with logging statements
  that record whether the vulnerability site was reached
  and whether the triggering condition was satisfied.
\end{itemize}
In our primary evaluation benchmark, Magma~\cite{hazimeh2020magma},
both prerequisites are natively satisfied.
\texttt{magma\_log} records reached and triggered states,
while ISAN mode promotes satisfied predicates to sanitizer-style crashes.
\sys additionally consumes reach/no-trigger diagnostics
and predicate semantics during generator synthesis.
While inferring bug predicate from source code or patch history is feasible for LLM-agents
(e.g., as shown in our 1-day vulnerability reproduction
experiment~\autoref{sec:eval:one-day-reproduction}),
we consider it as an orthogonal and out-of-scope problem for future work.

\begin{figure}[tbp]
  \centering
  \includegraphics[width=0.99\columnwidth,keepaspectratio]{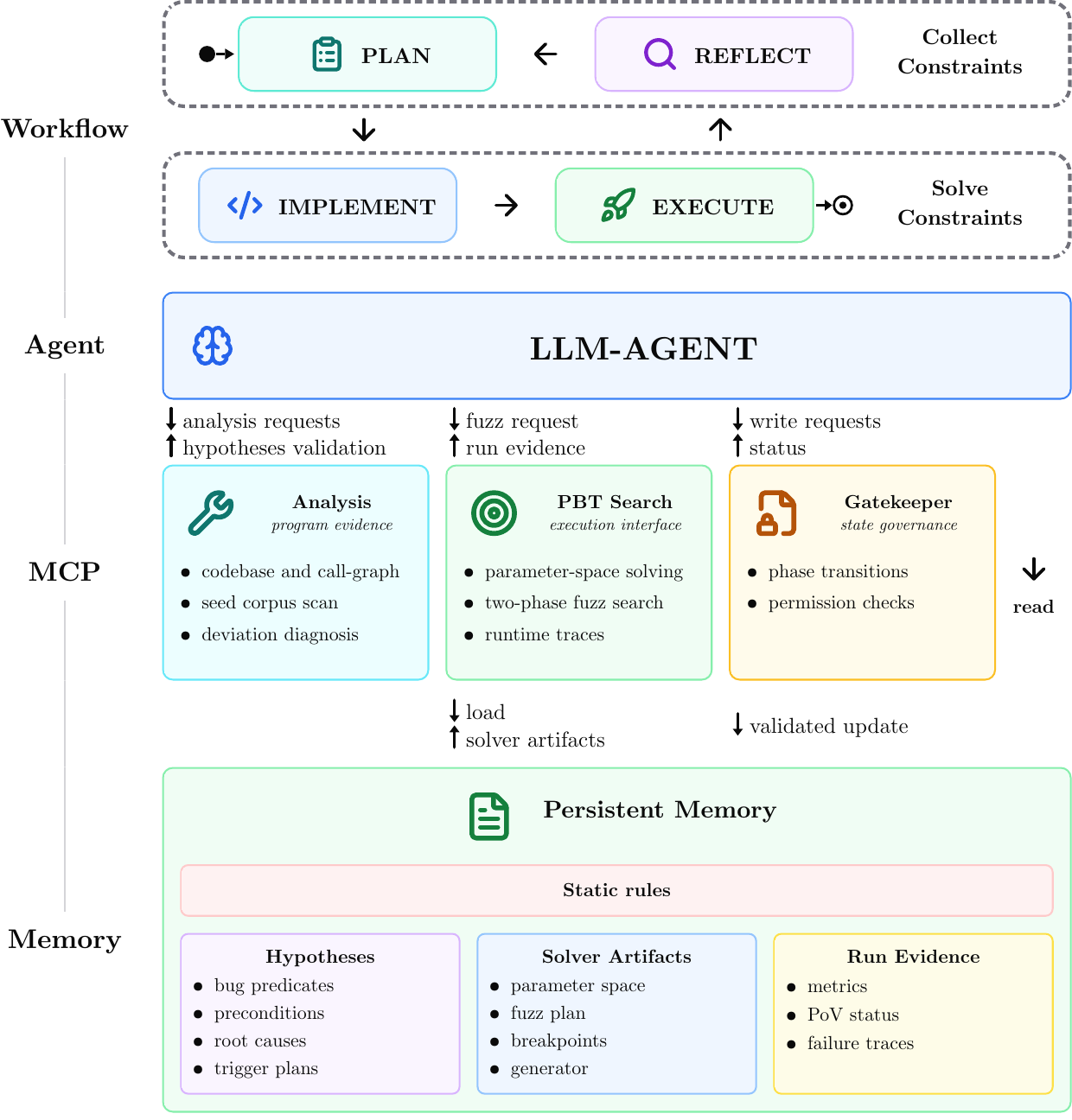}
  \caption{\sys Architecture.}
  \label{fig:arch}
  \vspace{-1em}
\end{figure}

\subsubsection{Architecture}
\sys employs a four-layer architecture (\autoref{fig:arch}) to enable
agentic property-based directed fuzzing.
At a high level, the \textbf{workflow layer} defines a state machine with four phases
forming a continuous \textbf{PLAN--IMPLEMENT--EXECUTE--REFLECT (PIER)} loop:
PLAN and REFLECT phases for inferring and refining semantic constraints
for reaching and triggering the vulnerability;
IMPLEMENT and EXECUTE for encoding the constraints into input-level parameter spaces
and efficient constraint solving.
Beneath it, the \textbf{LLM agent} acts as the ``brain'' to perform semantic reasoning
to (1) generate hypotheses about reaching and triggering constraints,
(2) infer root causes of vulnerabilities,
(3) formulate plausible ways to solve the constraints,
(4) synthesize the solver as a property-based fuzzer, and
(5) leverage runtime evidence to diagnose failures and refine hypotheses.
The \textbf{MCP tool layer} provides stateless tools to assist the agent's reasoning,
including call graph analysis, corpus analysis,
deviation detection, and a generic property-based fuzzing framework
with debugging feedback collection.
The \textbf{memory layer} maintains evolving hypothesis state in
a persistent Markdown file (\autoref{lst:workflow-state} in the appendix).
It prevents memory drift during extended reasoning tasks,
and redundant re-exploration of invalidated hypotheses.
A gatekeeper architecture mediates all state updates through three workflow MCP tools,
enforcing phase-specific permissions and preventing hallucination-induced drift.

The architecture of \sys combines the workflow pattern and
the agent pattern~\cite{Gulli2025AgenticDesignPatterns, anthropic2024agents}.
Purely workflow-driven systems lack adaptability
for unforeseeable vulnerability patterns and constraint discovery~\cite{anthropic2024agents}.
Conversely, unconstrained agents risk unbounded computational costs
and cascading reasoning errors from hallucination~\cite{zhu2025llm}.
Our hybrid approach constrains agent autonomy
within a workflow-defined state machine and corresponding state invariants,
enabling long-horizon reasoning while maintaining control.
We elaborate on each workflow phase
and state enforcement mechanisms in subsequent sections.

\subsubsection{Workflow}
The workflow starts with a system prompt (\autoref{lst:init-prompt} in the appendix)
that initializes the agent with task objectives (e.g., target vulnerability to trigger),
program context (e.g., how to test the target, signals for reaching and triggering),
and configuration details.
In each iteration,
the \textbf{PLAN} phase analyzes vulnerability sites and their corresponding dependencies
to formulate hypotheses about reachability and triggering constraints.
The \textbf{IMPLEMENT} phase synthesizes parameterized input generators
that encode program semantic-level constraints into input-level typed parameter spaces.
The \textbf{EXECUTE} phase performs two-phase PoV search with property-based fuzzing.
The fuzzer first evaluates a set of concrete parameters that represent
plausible solutions the agent believes will satisfy the constraints.
If phase one fails, the fuzzer systematically explores the constrained input spaces
via random sampling.
Upon discovering PoV, the workflow terminates;
otherwise, the \textbf{REFLECT} phase diagnoses failures (e.g., path divergence)
and refines hypothesis for the next PLAN iteration.

\subsection{PLAN Phase}

Building on the constraint taxonomy from \autoref{sec:motivation},
the LLM agent performs constraint inference
through autonomous dependency analysis to formulate hypotheses
about triggering the target vulnerability.
If this is not the first iteration, the agent leverages the failure diagnosis
results from the previous REFLECT phase to refine its hypotheses from the last PLAN phase.
Our prototype is built upon the cursor-cli agent~\cite{cursor2024},
which has built-in code exploration capabilities to perform backward slicing to
extract dependencies.
We provide four additional MCP tools to assist constraint inference:
\cc{get\_callers} and \cc{get\_callees} for call graph analysis,
\cc{get\_reaching\_routes} to retrieve reaching testcases,
and \cc{get\_corpus\_status} to monitor corpus processing.

The constraint inference workflow proceeds through four structured updates
to persistent memory in the markdown file.
These constraints form a constraint hierarchy.
First, the agent extracts vulnerability triggering predicates
as safety properties and stores in the \cc{BugPredicates} block
(e.g., the buffer overflow condition at \cc{valid.c:1342} for CVE-2017-9047).
We assume these predicates are provided beforehand;
for Magma~\cite{hazimeh2020magma}, we use predicates from \cc{MAGMA\_LOG} statements.

Second, starting from \cc{BugPredicates}, the agent performs autonomous backward
program slicing until the entry points, to infer the \emph{reachability constraints}.
The hypothesized constraints are stored in the \cc{\textbf{Preconditions}} block.
Four key tools are used to assist the agent's understanding and reasoning:
(1,2) the built-in \cc{Grep} and \cc{Glob} tool to search the codebase,
(3) the built-in \cc{Read} tool to read source code files, and
(4) the call graph MCP tool to resolve indirect call edges.
For CVE-2017-9047 in libxml2,
its \cc{Preconditions} block records five reachability constraints:
\begin{lstlisting}[style=pbfuzz-json]
[{"id": "R1", "statement": "Must reach xmlSnprintfElementContent function", "input_constraints": ["Input must be valid XML", "Must trigger DTD validation"]},
 {"id": "R2", "statement": "content->type must be XML_ELEMENT_CONTENT_ELEMENT", "input_constraints": ["DTD must contain ELEMENT declarations"]},
 {"id": "R3", "statement": "content->prefix must not be NULL", "input_constraints": ["Element must have namespace prefix in DTD"]},
 {"id": "R4", "statement": "Buffer size check at line 1302 must pass", "input_constraints": ["Initial buffer must have at least 50 bytes"]},
 {"id": "R5", "statement": "Buffer size check at line 1326 must pass", "input_constraints": ["In unfixed version, this check is disabled"]}]
\end{lstlisting}
Note that compared to traditional static backward slicing,
which can easily produce an overwhelming number of constraints and
reachable paths/basic blocks,
our agentic approach generates a substantially more concise set of semantic constraints,
owing to the LLM's code understanding and reasoning capabilities.

Third, the agent examines triggering constraints from \cc{BugPredicates} to
identify underlying vulnerability categories and stores them in the \cc{RootCauses} block
(e.g., buffer overflow for CVE-2017-9047).
Each \cc{RootCause} links to related \cc{Precondition} identifiers,
establishing which reachability constraints enable specific vulnerability triggers.
These root causes help the agent define parameter space boundaries during IMPLEMENT,
making generated inputs more likely to trigger the vulnerability.

Finally, the agent reasons about \cc{RootCauses} and \cc{Preconditions} to formulate
concrete and plausible vulnerability triggering strategies,
and stores the plans in the \cc{\textbf{TriggerPlans}} block.
Each \cc{TriggerPlan} references both reachability and triggering constraint identifiers.
It also contains a complexity score (1-10) based on constraint hierarchy difficulty,
and maintains execution status (\cc{pending}, \cc{in\_progress}, \cc{completed}, \cc{failed})
for future refinement.
For CVE-2017-9047, the plan is to create deeply nested DTD content models with
qualified names such that it consumes the 5000-byte buffer through recursive calls.
The corresponding \cc{TriggerPlans} records:
\begin{lstlisting}[style=pbfuzz-json]
[{"id": "TP1", "precondition_ids": ["R1", "R2", "R3", "R4"], "description": "Create DTD with nested element content using qualified names", "complexity": 3, "status": "pending"}]
\end{lstlisting}
Note that the agent can generate multiple \cc{TriggerPlans}.
Each plan targets a different combination of (1) a subset of \cc{Preconditions} and
(2) a single \cc{RootCause}.
During EXECUTE phases, our PBT fuzzer can explore several different \cc{TriggerPlans}
at the same time to improve the efficiency of PoV generation.

As mentioned earlier, if this is not the first PLAN iteration,
the agent leverages the failure diagnosis results from the previous REFLECT phase
to refine its hypotheses, which may involve:
(1) update \cc{Preconditions} to include identified missing constraints;
(2) revise \cc{RootCauses} to correct inaccurate understandings of the vulnerabilities; and
(3) abandon \cc{TriggerPlans} that are unlikely to succeed.

This constraint inference pipeline mimics human experts.
Starting with the target vulnerability,
the agent (1) examines the source code to understand the target code and
to extract dependencies; (2) uses dynamic information from corpus analysis
to validate reachability hypotheses (explained later); and (3) employs
call graph analysis to identify reaching execution paths.
Finally, the agent formulates triggering plans based on its understanding of
the vulnerability and the PUT.
If the hypotheses were wrong, the agent refines them based on failure diagnosis.

\subsection{IMPLEMENT Phase}

Based on constraint hierarchy from the PLAN phase, in the IMPLEMENT phase,
the agent reasons about how to \emph{translate} the high-level semantic constraints from
the \cc{TriggerPlans} into input-level parameter spaces.
Then, the agent synthesizes a custom solver (i.e., the PBT fuzzer) to search for
a (PoV) input that satisfies the identified reaching and triggering constraints.
We adopt this strategy instead of directly using LLMs to solve constraints
and generate inputs because recent research demonstrates that LLMs are ineffective at solving
complex arithmetic constraints~\cite{jiang2024towards},
and generating binary inputs~\cite{zhang2025low}.
Moreover, invoking LLMs for input generation is slow and costly.
Note that during the PLAN phase, the agent may generate multiple \cc{TriggerPlans}.
In this phase, we ask the agent to generate unified parameter spaces and
a single input generator that can cover all \cc{TriggerPlans}.
We provide two MCP tools to assist generator synthesis:
\cc{extract\_parameters} derives parameter specifications from reaching seeds in the corpus,
and \cc{get\_generator\_api\_doc} provides API documentation and examples.

Our custom solver consists of two key artifacts to realize a custom PBT fuzzer:
(1) a Python-based parameterized input generator,
and (2) a configuration file defining the parameter spaces,
concrete parameter sets, and breakpoint definitions.
These artifacts are generated in four steps.

First, the agent materializes input constraints from \cc{TriggerPlans} into
a typed parameter space specification,
and stores them in the \cc{\textbf{ParameterSpace}} section of the configuration file.
Each parameter is defined as a key-value pair,
where the keys are parameter space identifiers used by the input generator,
and the values define the parameter types
(\cc{int\_range}, \cc{float\_range}, \cc{categorical}, \cc{segments}, \cc{base\_seed})
and value domains (e.g., min/max ranges for numerical types,
allowed values for categorical types).
In this step, the agent may invoke the \cc{extract\_parameters} tool to retrieve
parameter domains from known reaching testcases from the corpus to help
reason about value ranges, categorical enumerates, and boundary values
that may satisfy the constraint hierarchy extracted in PLAN phase.
For CVE-2017-9047,
its \cc{ParameterSpace} section records:
\begin{lstlisting}[style=pbfuzz-json]
{"element_prefix_length": {"type": "int_range", "min": 1, "max": 200},
 "element_name_length": {"type": "int_range", "min": 1, "max": 200},
 "nesting_depth": {"type": "int_range", "min": 1, "max": 50},
 "num_elements": {"type": "int_range", "min": 1, "max": 100},
 "content_model_type": {"type": "categorical", "values": ["SEQ", "OR", "MIXED"]}}
\end{lstlisting}
Second, the agent synthesizes a parameterized input generator
implementing a generic \cc{generate(**params) -> bytes} Python interface function
that maps parameter tuples to concrete test inputs.
For CVE-2017-9047, the generator creates nested DTD structures
mirroring the recursive control flow identified in~\autoref{sec:motivation}.
Third, to improve search efficiency, the agent generates 5-10 concrete parameter sets
stored in \cc{ConcreteParameters}, prioritizing malformed or boundary inputs
(e.g., for CVE-2017-9047: \cc{depth=10, names=70}).
Fourth, the agent specifies debugging breakpoints in \cc{Breakpoints}
to collect runtime evidence for hypothesis validation
(e.g., capturing \cc{size}, \cc{len}, and buffer consumption at line 1342).

In summary, the IMPLEMENT phase realizes our key insight:
\emph{property-based testing (PBT) provides
the missing bridge between semantic analysis and constraint solving}
It translates high-level semantic constraint
hierarchy from the PLAN phase (\cc{Preconditions}, \cc{RootCauses}, \cc{TriggerPlans})
into input space executable artifacts
(\cc{ParameterSpace}, \cc{generate}, \cc{ConcreteParameters}, \cc{Breakpoints})
that guide the EXECUTE phase to search for the PoV input.

\begin{figure}[tbp]
  \centering
  \includegraphics[width=0.7\columnwidth]{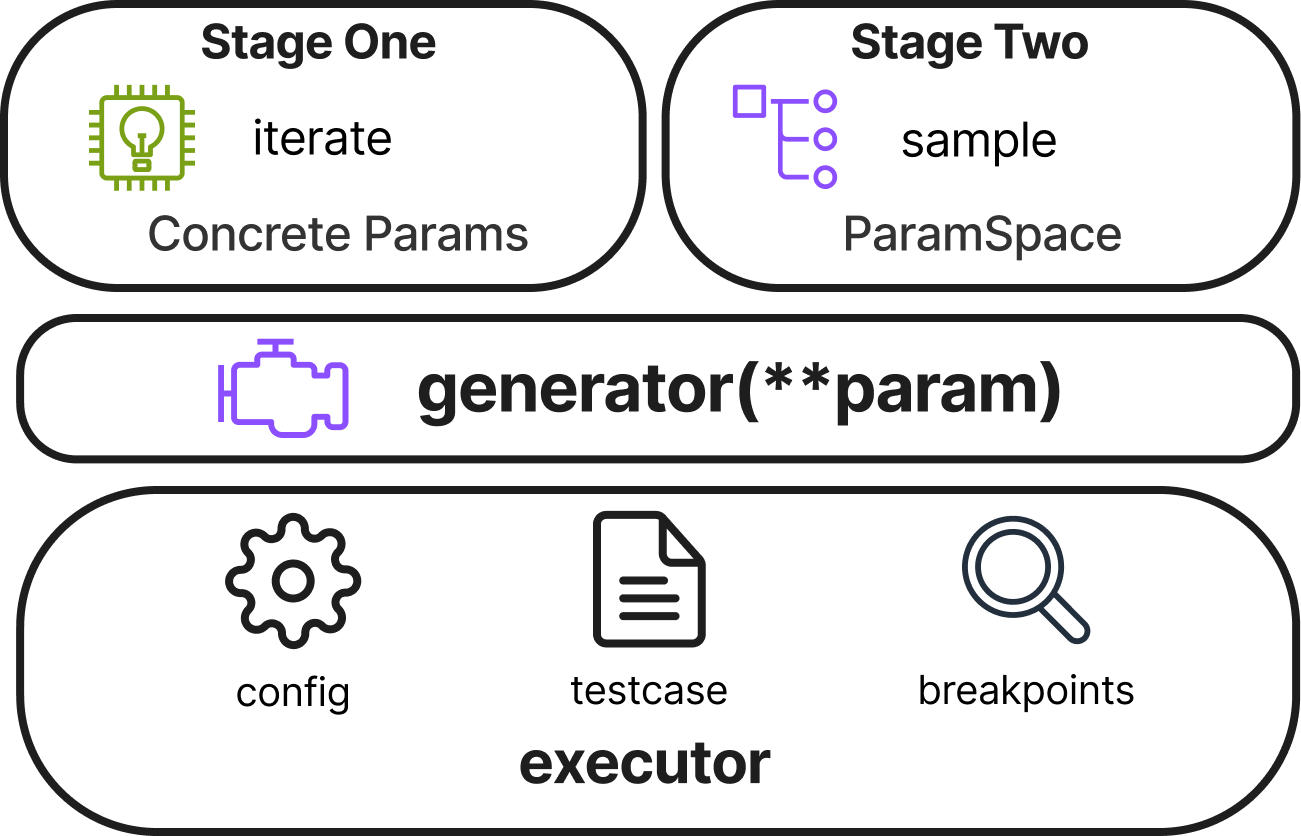}
  \caption{
  Property-based fuzzing with two-stage test generation:
  Stage 1 iterates concrete parameters,
  Stage 2 performs heuristic parameter space sampling.}
  \label{fig:pbfuzz}
  \vspace{-1em}
\end{figure}

\subsection{EXECUTE and REFLECT Phases}

With executable artifacts ready, the agent invokes the \cc{fuzz} MCP tool 
to perform two-stage PoV search via property-based testing.
As shown in \autoref{fig:pbfuzz}, the fuzzer performs two-stage search.
Stage 1 evaluates concrete parameters from \cc{ConcreteParameters},
testing agent-hypothesized solutions with debugger inspection at \cc{Breakpoints}.
If unsuccessful, Stage 2 performs systematic sampling from the \cc{ParameterSpace},
prioritizing boundary values and constraint edges
that are more likely to trigger vulnerabilities~\cite{claessen2011quickcheck}.

Upon discovering PoV, the workflow terminates with success.
For CVE-2017-9047, the fuzzer discovers triggering inputs at iteration 156
with parameters \cc{depth=10}, \cc{prefix/name=70}.

If PoV discovery fails, the REFLECT phase diagnoses failures
using two MCP tools:
\cc{detect\_deviation} identifies violated reachability constraints,
and \cc{launch\_gdb} enables interactive debugging for triggering constraint analysis.

Failure diagnosis distinguishes reachability failures from triggering failures.
For non-reaching testcases, \cc{detect\_deviation} identifies execution divergence points
by comparing against pre-computed critical locations
(boundaries between reachable and unreachable regions).
For reach-but-no-trigger testcases, the agent performs backward dependency analysis
to identify violated triggering constraints.
Diagnostic evidence guides the next PLAN iteration to refine
\cc{Preconditions}, \cc{RootCauses}, and \cc{TriggerPlans}.
This cycle continues until PoV is discovered or resource limits are exhausted.

\subsection{Implementation Details}

We now detail the concrete realization
of our architecture across three infrastructure components.

\PP{Agent Infrastructure}
We implement the agent layer using cursor-cli~\cite{cursor2024},
which provides built-in semantic search and grep tools
for efficient codebase navigation.
The system invokes the agent via shell interface,
passing system prompts through stdin.

\PP{MCP Tool Implementation}
We implement six MCP servers with 7,458 lines of Python code.
They communicate with cursor-cli via JSON-RPC over standard I/O.
The fuzzer tool dynamically loads and executes Python generator code
in an isolated namespace at runtime.

\PP{Static Analysis}
For static analysis, we first compile the target project
using \cc{clang} with link-time optimization (LTO) enabled.
This allows us to reuse existing build systems to generate LLVM bitcode.
We then perform whole-program static analysis based on the pre-opt bitcode
dumped by \cc{lld} during linking.
We use a type-based call graph analysis~\cite{lu2016unisan}
to pre-compute call graphs (997 lines of C++).
Although imprecise, this analysis is much faster and we believe LLM agents can
overcome the imprecision through source code reasoning.
To identify critical locations for deviation detection, we implement
a backward breadth-first search (1,446 lines of C++) to mark (context-insensitive)
basic blocks that (1) can reach the target locations,
and (2) can reach known exit points
(e.g., \cc{exit()} syscall, return of \cc{main()}).
We perform the static analysis before the workflow starts,
and save the results to files.
MCP tools \cc{get\_callers/get\_callees} and \cc{detect\_deviation}
directly query these pre-computed results,
enabling efficient reachability constraint collection during PLAN
and failure diagnosis during REFLECT.

\section{Evaluation}
\label{sec:eval}

We evaluate \sys to validate whether the agentic fuzzing architecture
enables effective PoV generation through iterative constraint inference and solving.
Our evaluation addresses two fundamental hypotheses:
(H1) agents can successfully infer reachability and triggering constraints,
with the help of MCP tools, and map these constraints to input parameter spaces;
and (H2) property-based testing can solve extracted constraints efficiently.
We validate our hypotheses by systematically investigating the following research questions:

\begin{enumerate}[label=\textbf{RQ\arabic*.}]
  \item \textbf{Overall Effectiveness}: Can our agentic approach successfully
  generate proof-of-vulnerability (PoV) inputs against CVEs in real-world benchmarks,
  especially those that challenge existing techniques like directed fuzzers?

  \item \textbf{Strengths and Limitations}: Through empirical case studies,
  we evaluate our hypotheses in-depth and identify
  what unique strengths and limitations
  our approach exhibits compared to existing PoV generation techniques?

  \item \textbf{Ablation Study}: What components contribute to system effectiveness?

  \item \textbf{New Bug Discovery}: Can our approach discover new vulnerabilities?
\end{enumerate}

\subsection{Experimental Setup}

We evaluate \sys on the Magma benchmark~\cite{hazimeh2020magma},
a ground-truth fuzzing suite with real-world CVEs.
We compare against directed greybox fuzzers (AFLGo, SelectFuzz),
a coverage-guided fuzzer (AFL++), an LLM-assisted fuzzer (G2Fuzz),
and nine recent fuzzers from literature (RQ1).
Complete benchmarking details, hardware configurations, baseline versions,
and baseline parameters are documented in \autoref{sec:open-science}
to facilitate reproducibility and artifact review.
For fuzzers, we ran 10 trials of 24 hours each per target, following Magma's protocol.
Due to cost constraints, however, we only ran \sys for a single trial of 30 minutes per target.

\subsubsection{Metrics}

We measure PoV generation systems with two primary metrics:
\begin{itemize}[leftmargin=*]
\item \textbf{CVE coverage} quantifies the number of unique vulnerabilities triggered.
    This is the effectiveness metric used in RQ1.
    By comparing vulnerabilities that can only be triggered by a system,
    we can also understand the unique strengths of different approaches.
\item \textbf{Time-to-exposure (TTE)} records elapsed time until first PoV generation.
    This metric captures the efficiency of a PoV generation approach.
\end{itemize}
We also report Time-to-reach (TTR), the elapsed time until the first input reaches
the vulnerable code location and a metric the Magma benchmark reports,
but we found it less informative than TTE since 60 vulnerabilities are reachable
with the initial seed inputs.
For vulnerabilities triggerable from multiple fuzzing harnesses,
we report the optimal TTE and TTR.
We also measure \sys's efficiency via (1) PIER iteration count until PoV discovery
and (2) LLM API cost in total tokens consumed,
supplemented by qualitative case studies of constraint hierarchies and failure modes.

\begin{figure}[hbtp]
\centering
\includegraphics[width=\columnwidth]{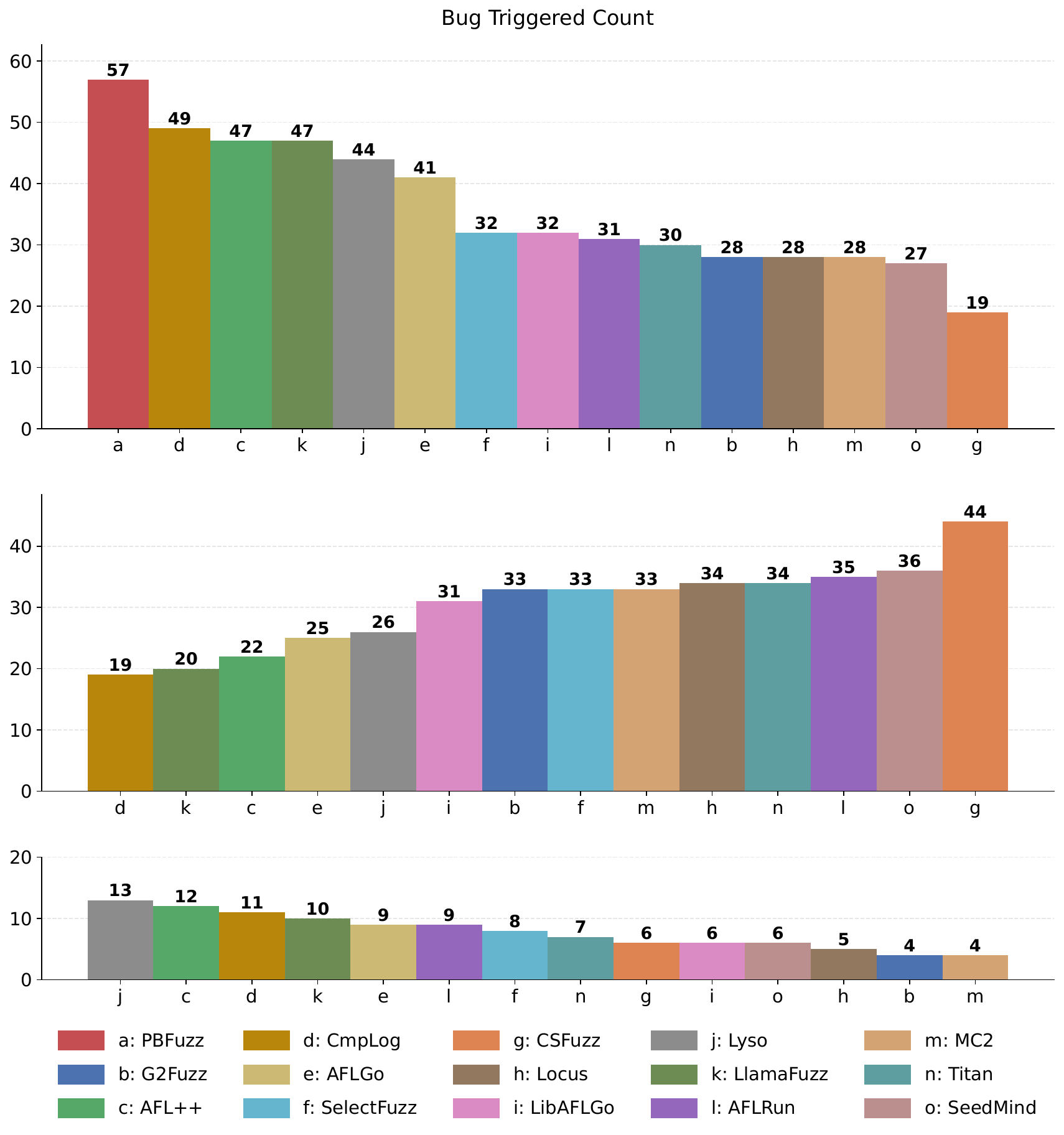}
\caption{%
  Effectiveness across Magma benchmarks ordered by performance.
  Top: total unique vulnerabilities triggered.
  Middle: number of vulnerabilities exclusively triggered by \sys compared to other fuzzers.
  Bottom: compared to \sys, vulnerabilities exclusively discovered by baselines.
}
\label{fig:fuzzer-comparison}
\vspace{-1em}
\end{figure}

\subsection{RQ1. Overall Effectiveness}
\label{sec:eval:effectiveness}

\subsubsection{CVE Coverage}
\autoref{fig:fuzzer-comparison} summarizes PoV generation effectiveness
across all evaluated fuzzers on Magma, where we union the CVEs triggered across
all trials of fuzzing campaigns.
Overall, \sys triggered \textbf{57} vulnerabilities across Magma targets,
and outperformed all baseline fuzzers (first row).
Upon inspection, we found that 6 vulnerabilities uniquely triggered by \sys
(\cc{PHP001}, \cc{PHP003}, \cc{PHP010}, \cc{PHP013}, \cc{PHP014}, \cc{SSL006})
were due to the use of custom harnesses
rather than improved PoV generation on the same harnesses
(more details in~\autoref{sec:eval:strengths-limitations}).
However, even after excluding these 6 harness-specific cases,
\sys still triggered 51 vulnerabilities, outperforming AFL++ with CmpLog (49)
and validating core PoV generation capability independent of harness adaptation.
It also outperformed all directed greybox fuzzers and LLM-assisted fuzzers.
Notably, Llamafuzz executed for a month, while other baseline fuzzers were run
for 10 trials of 24 hours each.
In contrast, \sys only ran for 30 minutes in a single trial.

\subsubsection{Unique CVE Coverage}
The second row of \autoref{fig:fuzzer-comparison} shows the number of vulnerabilities
that were uniquely triggered by \sys compared to each baseline;
and the third row shows the number of vulnerabilities uniquely triggered by each baseline
compared to \sys.
Overall, \sys consistently triggers unique vulnerabilities
that none of the baselines can trigger.
\sys triggered {\bf 17} vulnerabilities that none of the other fuzzers could trigger:
\cc{LUA001}, \cc{PDF002}, \cc{PDF005}, \cc{PDF008}, \cc{PDF009}, \cc{PDF012}, \cc{PDF022}, \cc{SND016}, \cc{SQL007}, \cc{SQL010}, \cc{XML010}
with the same harnesses as other fuzzers; 
and the aforementioned 6 harness-specific vulnerabilities.
Among different fuzzers,
\sys uniquely triggered 19 vulnerabilities over AFL++ with CmpLog,
and 25 over AFLGo;
while AFL++-cmplog uniquely triggered 11 vulnerabilities over \sys,
and AFLGo uniquely triggered 9 vulnerabilities over \sys.
This result shows that \sys provides unique strengths in PoV generation capabilities
that complement existing fuzzing techniques.
We provide in-depth analysis of how \sys triggered these unique vulnerabilities in RQ2.

\subsubsection{Efficiency (Time-to-Exposure)}
\autoref{tab:trigger} details the Time-to-Exposure (TTE) statistics of different approaches.
\sys's min/median/max TTE is 83s/339s/1,441s,
compared to 5s/8,680s/85,322s for AFL++ with CmpLog,
3s/6,170s/85,200s for AFLGo, 2s/3,397s/80,394s for SelectFuzz, and 6s/4,751s/43,803s for G2Fuzz.
Evidently, \sys not only triggered more vulnerabilities, including more unique vulnerabilities,
but also did so in a single iteration within a \emph{30-minute} budget.

This shows that our PLAN-IMPLEMENT-EXECUTE workflow generates PoVs far more
efficiently than conventional approaches, including LLM-assisted fuzzers like G2Fuzz,
despite relying on LLMs.
This performance gap aligns with the theoretical ``efficiency boundary'' of
random testing~\cite{bohme2014efficiency,bohme2015probabilistic},
where the discovery rate of greybox fuzzers decays exponentially.
Consequently, simply extending the time budget---as seen with Llamafuzz's one-month campaign---yields diminishing returns.
\sys overcomes this limitation by adopting semantic reasoning and property-based testing.
Leveraging the ``asymmetry of verification''~\cite{wei2024asymmetry} inherent in PoV tasks
(which are hard to generate but easy to verify),
\sys utilizes generative AI to infer semantic constraints, then generate a custom solver.
This shifts the computational burden from inefficient random sampling to targeted solving,
unleashing real \emph{directed fuzzing}.

\begin{table}[!htbp]
\caption{
Time-to-Exposure (TTE) comparison across benchmarks.
T.O denotes a 24-hour timeout; T.O* denotes \sys's 30-minute budget exhausted.
NA denotes the fuzzer cannot handle the target.
}
\label{tab:trigger}
{\footnotesize
\begin{tabular}{l|r rr rr rr rr}
\toprule
\textbf{Bug ID} & \textbf{PBFuzz} & \textbf{G2Fuzz} & \multicolumn{1}{c}{\textbf{AFL++}} & \multicolumn{1}{c}{\textbf{cmplog}} & \multicolumn{1}{c}{\textbf{AFLGo}} & \multicolumn{1}{c}{\textbf{SelectFuzz}} \\
\midrule
LUA001 & \cellcolor[HTML]{165128} 2.1m & \cellcolor[HTML]{FF9999} T.O & \cellcolor[HTML]{FF9999} T.O & \cellcolor[HTML]{FF9999} T.O & \cellcolor[HTML]{FF9999} T.O & \cellcolor[HTML]{FF9999} T.O \\
LUA002 & \cellcolor[HTML]{165128} 7.2m & \cellcolor[HTML]{FF9999} T.O & \cellcolor[HTML]{60b472} 22.7h & \cellcolor[HTML]{8ac794} 23.7h & \cellcolor[HTML]{FF9999} T.O & \cellcolor[HTML]{FF9999} T.O \\
LUA003 & \cellcolor[HTML]{165128} 5.5m & \cellcolor[HTML]{FF9999} T.O & \cellcolor[HTML]{FF9999} T.O & \cellcolor[HTML]{FF9999} T.O & \cellcolor[HTML]{FF9999} T.O & \cellcolor[HTML]{FF9999} T.O \\
LUA004 & \cellcolor[HTML]{60b472} 4m & \cellcolor[HTML]{165128} 6s & \cellcolor[HTML]{d8edda} 11.3h & \cellcolor[HTML]{ffffff} 18.1h & \cellcolor[HTML]{b2dab7} 8.8h & \cellcolor[HTML]{8ac794} 4.7h \\
PDF002 & \cellcolor[HTML]{165128} 2.9m & \cellcolor[HTML]{FF9999} T.O & \cellcolor[HTML]{FF9999} T.O & \cellcolor[HTML]{FF9999} T.O & \cellcolor[HTML]{FF9999} T.O & \cellcolor[HTML]{FF9999} T.O \\
PDF003 & \cellcolor[HTML]{165128} 5.3m & \cellcolor[HTML]{60b472} 5.2h & \cellcolor[HTML]{d8edda} 6.7h & \cellcolor[HTML]{ffffff} 8.7h & \cellcolor[HTML]{b2dab7} 6.2h & \cellcolor[HTML]{8ac794} 5.3h \\
PDF005 & \cellcolor[HTML]{165128} 12.4m & \cellcolor[HTML]{FF9999} T.O & \cellcolor[HTML]{FF9999} T.O & \cellcolor[HTML]{FF9999} T.O & \cellcolor[HTML]{FF9999} T.O & \cellcolor[HTML]{FF9999} T.O \\
PDF006 & \cellcolor[HTML]{FF9999} T.O* & \cellcolor[HTML]{FF9999} T.O & \cellcolor[HTML]{165128} 20.3h & \cellcolor[HTML]{FF9999} T.O & \cellcolor[HTML]{FF9999} T.O & \cellcolor[HTML]{FF9999} T.O \\
PDF008 & \cellcolor[HTML]{165128} 3.2m & \cellcolor[HTML]{FF9999} T.O & \cellcolor[HTML]{FF9999} T.O & \cellcolor[HTML]{FF9999} T.O & \cellcolor[HTML]{FF9999} T.O & \cellcolor[HTML]{FF9999} T.O \\
PDF009 & \cellcolor[HTML]{165128} 3.1m & \cellcolor[HTML]{FF9999} T.O & \cellcolor[HTML]{FF9999} T.O & \cellcolor[HTML]{FF9999} T.O & \cellcolor[HTML]{FF9999} T.O & \cellcolor[HTML]{FF9999} T.O \\
PDF010 & \cellcolor[HTML]{165128} 8.4m & \cellcolor[HTML]{b2dab7} 4.1h & \cellcolor[HTML]{d8edda} 5h & \cellcolor[HTML]{ffffff} 5.2h & \cellcolor[HTML]{8ac794} 20.9m & \cellcolor[HTML]{60b472} 8.6m \\
PDF011 & \cellcolor[HTML]{FF9999} T.O* & \cellcolor[HTML]{165128} 6.6h & \cellcolor[HTML]{8ac794} 18.8h & \cellcolor[HTML]{60b472} 17.7h & \cellcolor[HTML]{FF9999} T.O & \cellcolor[HTML]{b2dab7} 21.8h \\
PDF012 & \cellcolor[HTML]{165128} 5.6m & \cellcolor[HTML]{FF9999} T.O & \cellcolor[HTML]{FF9999} T.O & \cellcolor[HTML]{FF9999} T.O & \cellcolor[HTML]{FF9999} T.O & \cellcolor[HTML]{FF9999} T.O \\
PDF016 & \cellcolor[HTML]{d8edda} 3.2m & \cellcolor[HTML]{ffffff} 10.6m & \cellcolor[HTML]{165128} 19s & \cellcolor[HTML]{60b472} 29s & \cellcolor[HTML]{8ac794} 69s & \cellcolor[HTML]{b2dab7} 87s \\
PDF018 & \cellcolor[HTML]{165128} 3.1m & \cellcolor[HTML]{b2dab7} 9h & \cellcolor[HTML]{d8edda} 10.7h & \cellcolor[HTML]{ffffff} 22.5h & \cellcolor[HTML]{8ac794} 21.9m & \cellcolor[HTML]{60b472} 10.6m \\
PDF019 & \cellcolor[HTML]{165128} 22.9m & \cellcolor[HTML]{FF9999} T.O & \cellcolor[HTML]{60b472} 2h & \cellcolor[HTML]{8ac794} 6h & \cellcolor[HTML]{FF9999} T.O & \cellcolor[HTML]{FF9999} T.O \\
PDF021 & \cellcolor[HTML]{165128} 8.1m & \cellcolor[HTML]{FF9999} T.O & \cellcolor[HTML]{FF9999} T.O & \cellcolor[HTML]{60b472} 21.2h & \cellcolor[HTML]{b2dab7} 23.5h & \cellcolor[HTML]{8ac794} 22.3h \\
PDF022 & \cellcolor[HTML]{165128} 3.1m & \cellcolor[HTML]{FF9999} T.O & \cellcolor[HTML]{FF9999} T.O & \cellcolor[HTML]{FF9999} T.O & \cellcolor[HTML]{FF9999} T.O & \cellcolor[HTML]{FF9999} T.O \\
PHP001 & \cellcolor[HTML]{165128} 11m & \cellcolor[HTML]{FF9999} T.O & \cellcolor[HTML]{FF9999} T.O & \cellcolor[HTML]{FF9999} T.O & \cellcolor[HTML]{FF9999} T.O & \cellcolor[HTML]{FF9999} NA \\
PHP003 & \cellcolor[HTML]{165128} 12.3m & \cellcolor[HTML]{FF9999} T.O & \cellcolor[HTML]{FF9999} T.O & \cellcolor[HTML]{FF9999} T.O & \cellcolor[HTML]{FF9999} T.O & \cellcolor[HTML]{FF9999} NA \\
PHP004 & \cellcolor[HTML]{60b472} 10m & \cellcolor[HTML]{165128} 4.1m & \cellcolor[HTML]{8ac794} 43.5m & \cellcolor[HTML]{b2dab7} 91.5m & \cellcolor[HTML]{d8edda} 2.3h & \cellcolor[HTML]{FF9999} NA \\
PHP009 & \cellcolor[HTML]{8ac794} 15.4m & \cellcolor[HTML]{60b472} 5m & \cellcolor[HTML]{165128} 3.6m & \cellcolor[HTML]{b2dab7} 17.7m & \cellcolor[HTML]{d8edda} 74m & \cellcolor[HTML]{FF9999} NA \\
PHP010 & \cellcolor[HTML]{165128} 19.6m & \cellcolor[HTML]{FF9999} T.O & \cellcolor[HTML]{FF9999} T.O & \cellcolor[HTML]{FF9999} T.O & \cellcolor[HTML]{FF9999} T.O & \cellcolor[HTML]{FF9999} NA \\
PHP011 & \cellcolor[HTML]{d8edda} 10.2m & \cellcolor[HTML]{60b472} 21s & \cellcolor[HTML]{b2dab7} 26s & \cellcolor[HTML]{8ac794} 23s & \cellcolor[HTML]{165128} 14s & \cellcolor[HTML]{FF9999} NA \\
PHP013 & \cellcolor[HTML]{165128} 6.7m & \cellcolor[HTML]{FF9999} T.O & \cellcolor[HTML]{FF9999} T.O & \cellcolor[HTML]{FF9999} T.O & \cellcolor[HTML]{FF9999} T.O & \cellcolor[HTML]{FF9999} NA \\
PHP014 & \cellcolor[HTML]{165128} 3.9m & \cellcolor[HTML]{FF9999} T.O & \cellcolor[HTML]{FF9999} T.O & \cellcolor[HTML]{FF9999} T.O & \cellcolor[HTML]{FF9999} T.O & \cellcolor[HTML]{FF9999} NA \\
PNG001 & \cellcolor[HTML]{165128} 10.8m & \cellcolor[HTML]{FF9999} T.O & \cellcolor[HTML]{60b472} 17.4h & \cellcolor[HTML]{8ac794} 22.5h & \cellcolor[HTML]{FF9999} T.O & \cellcolor[HTML]{FF9999} T.O \\
PNG003 & \cellcolor[HTML]{ffffff} 6.7m & \cellcolor[HTML]{d8edda} 2.9m & \cellcolor[HTML]{8ac794} 4s & \cellcolor[HTML]{b2dab7} 5s & \cellcolor[HTML]{60b472} 3s & \cellcolor[HTML]{165128} 2s \\
PNG006 & \cellcolor[HTML]{8ac794} 4.5m & \cellcolor[HTML]{60b472} 2.8m & \cellcolor[HTML]{FF9999} T.O & \cellcolor[HTML]{165128} 8s & \cellcolor[HTML]{FF9999} T.O & \cellcolor[HTML]{FF9999} T.O \\
PNG007 & \cellcolor[HTML]{165128} 11.9m & \cellcolor[HTML]{FF9999} T.O & \cellcolor[HTML]{8ac794} 72.2m & \cellcolor[HTML]{b2dab7} 1.8h & \cellcolor[HTML]{d8edda} 3.8h & \cellcolor[HTML]{60b472} 44.3m \\
SND001 & \cellcolor[HTML]{165128} 83s & \cellcolor[HTML]{b2dab7} 18.4m & \cellcolor[HTML]{d8edda} 18.6m & \cellcolor[HTML]{ffffff} 19.8m & \cellcolor[HTML]{8ac794} 5.3m & \cellcolor[HTML]{60b472} 83s \\
SND005 & \cellcolor[HTML]{8ac794} 4m & \cellcolor[HTML]{ffffff} 9.1h & \cellcolor[HTML]{b2dab7} 71.2m & \cellcolor[HTML]{d8edda} 94.7m & \cellcolor[HTML]{60b472} 34s & \cellcolor[HTML]{165128} 17s \\
SND006 & \cellcolor[HTML]{FF9999} T.O* & \cellcolor[HTML]{FF9999} T.O & \cellcolor[HTML]{8ac794} 77.8m & \cellcolor[HTML]{b2dab7} 99m & \cellcolor[HTML]{60b472} 6.8m & \cellcolor[HTML]{165128} 4.4m \\
SND007 & \cellcolor[HTML]{FF9999} T.O* & \cellcolor[HTML]{FF9999} T.O & \cellcolor[HTML]{b2dab7} 2.6h & \cellcolor[HTML]{8ac794} 2.1h & \cellcolor[HTML]{60b472} 6.7m & \cellcolor[HTML]{165128} 1.9m \\
SND016 & \cellcolor[HTML]{165128} 6.4m & \cellcolor[HTML]{FF9999} T.O & \cellcolor[HTML]{FF9999} T.O & \cellcolor[HTML]{FF9999} T.O & \cellcolor[HTML]{FF9999} T.O & \cellcolor[HTML]{FF9999} T.O \\
SND017 & \cellcolor[HTML]{60b472} 3.3m & \cellcolor[HTML]{165128} 30s & \cellcolor[HTML]{d8edda} 97.3m & \cellcolor[HTML]{ffffff} 1.7h & \cellcolor[HTML]{8ac794} 27.7m & \cellcolor[HTML]{b2dab7} 60.2m \\
SND020 & \cellcolor[HTML]{165128} 2m & \cellcolor[HTML]{ffffff} 4.8h & \cellcolor[HTML]{b2dab7} 1.9h & \cellcolor[HTML]{d8edda} 2.5h & \cellcolor[HTML]{8ac794} 1.8h & \cellcolor[HTML]{60b472} 1.7h \\
SND024 & \cellcolor[HTML]{FF9999} T.O* & \cellcolor[HTML]{FF9999} T.O & \cellcolor[HTML]{8ac794} 59.2m & \cellcolor[HTML]{b2dab7} 99.3m & \cellcolor[HTML]{60b472} 6.6m & \cellcolor[HTML]{165128} 1.8m \\
SQL002 & \cellcolor[HTML]{FF9999} T.O* & \cellcolor[HTML]{d8edda} 6.8h & \cellcolor[HTML]{165128} 5.1m & \cellcolor[HTML]{60b472} 9.5m & \cellcolor[HTML]{b2dab7} 55.6m & \cellcolor[HTML]{8ac794} 16.9m \\
SQL003 & \cellcolor[HTML]{165128} 2.4m & \cellcolor[HTML]{FF9999} T.O & \cellcolor[HTML]{FF9999} T.O & \cellcolor[HTML]{FF9999} T.O & \cellcolor[HTML]{60b472} 23.7h & \cellcolor[HTML]{FF9999} T.O \\
SQL007 & \cellcolor[HTML]{165128} 5.7m & \cellcolor[HTML]{FF9999} T.O & \cellcolor[HTML]{FF9999} T.O & \cellcolor[HTML]{FF9999} T.O & \cellcolor[HTML]{FF9999} T.O & \cellcolor[HTML]{FF9999} T.O \\
SQL010 & \cellcolor[HTML]{165128} 5.6m & \cellcolor[HTML]{FF9999} T.O & \cellcolor[HTML]{FF9999} T.O & \cellcolor[HTML]{FF9999} T.O & \cellcolor[HTML]{FF9999} T.O & \cellcolor[HTML]{FF9999} T.O \\
SQL012 & \cellcolor[HTML]{165128} 2m & \cellcolor[HTML]{FF9999} T.O & \cellcolor[HTML]{8ac794} 20.2h & \cellcolor[HTML]{d8edda} 22.5h & \cellcolor[HTML]{60b472} 16.2h & \cellcolor[HTML]{b2dab7} 21.2h \\
SQL013 & \cellcolor[HTML]{FF9999} T.O* & \cellcolor[HTML]{FF9999} T.O & \cellcolor[HTML]{FF9999} T.O & \cellcolor[HTML]{165128} 21.9h & \cellcolor[HTML]{FF9999} T.O & \cellcolor[HTML]{FF9999} T.O \\
SQL014 & \cellcolor[HTML]{FF9999} T.O* & \cellcolor[HTML]{d8edda} 12.2h & \cellcolor[HTML]{165128} 47.1m & \cellcolor[HTML]{60b472} 2.4h & \cellcolor[HTML]{b2dab7} 9.1h & \cellcolor[HTML]{8ac794} 3.3h \\
SQL015 & \cellcolor[HTML]{165128} 8.2m & \cellcolor[HTML]{FF9999} T.O & \cellcolor[HTML]{8ac794} 20.6h & \cellcolor[HTML]{60b472} 20.6h & \cellcolor[HTML]{FF9999} T.O & \cellcolor[HTML]{FF9999} T.O \\
SQL018 & \cellcolor[HTML]{165128} 3.5m & \cellcolor[HTML]{ffffff} 6.8h & \cellcolor[HTML]{60b472} 15.7m & \cellcolor[HTML]{b2dab7} 53.8m & \cellcolor[HTML]{d8edda} 60.1m & \cellcolor[HTML]{8ac794} 34.8m \\
SQL020 & \cellcolor[HTML]{165128} 2.1m & \cellcolor[HTML]{FF9999} T.O & \cellcolor[HTML]{60b472} 11.5h & \cellcolor[HTML]{8ac794} 17.3h & \cellcolor[HTML]{b2dab7} 17.8h & \cellcolor[HTML]{d8edda} 17.8h \\
SSL001 & \cellcolor[HTML]{FF9999} T.O* & \cellcolor[HTML]{FF9999} T.O & \cellcolor[HTML]{165128} 2.7h & \cellcolor[HTML]{60b472} 8.3h & \cellcolor[HTML]{8ac794} 10.9h & \cellcolor[HTML]{FF9999} NA \\
SSL002 & \cellcolor[HTML]{b2dab7} 4m & \cellcolor[HTML]{d8edda} 47.9m & \cellcolor[HTML]{165128} 94s & \cellcolor[HTML]{8ac794} 2.4m & \cellcolor[HTML]{60b472} 1.8m & \cellcolor[HTML]{FF9999} NA \\
SSL003 & \cellcolor[HTML]{b2dab7} 7.7m & \cellcolor[HTML]{d8edda} 47.2m & \cellcolor[HTML]{165128} 86s & \cellcolor[HTML]{8ac794} 2.6m & \cellcolor[HTML]{60b472} 1.9m & \cellcolor[HTML]{FF9999} NA \\
SSL006 & \cellcolor[HTML]{165128} 24m & \cellcolor[HTML]{FF9999} T.O & \cellcolor[HTML]{FF9999} T.O & \cellcolor[HTML]{FF9999} T.O & \cellcolor[HTML]{FF9999} T.O & \cellcolor[HTML]{FF9999} NA \\
SSL009 & \cellcolor[HTML]{FF9999} T.O* & \cellcolor[HTML]{165128} 76m & \cellcolor[HTML]{60b472} 14.4h & \cellcolor[HTML]{b2dab7} 20h & \cellcolor[HTML]{8ac794} 15.6h & \cellcolor[HTML]{FF9999} NA \\
SSL020 & \cellcolor[HTML]{165128} 9.1m & \cellcolor[HTML]{FF9999} T.O & \cellcolor[HTML]{60b472} 12.6h & \cellcolor[HTML]{b2dab7} 21.7h & \cellcolor[HTML]{8ac794} 20.7h & \cellcolor[HTML]{FF9999} NA \\
TIF001 & \cellcolor[HTML]{FF9999} T.O* & \cellcolor[HTML]{FF9999} T.O & \cellcolor[HTML]{60b472} 23h & \cellcolor[HTML]{165128} 22.5h & \cellcolor[HTML]{FF9999} T.O & \cellcolor[HTML]{FF9999} T.O \\
TIF002 & \cellcolor[HTML]{FF9999} T.O* & \cellcolor[HTML]{FF9999} T.O & \cellcolor[HTML]{165128} 5.8h & \cellcolor[HTML]{60b472} 8.9h & \cellcolor[HTML]{8ac794} 18.9h & \cellcolor[HTML]{b2dab7} 21.7h \\
TIF005 & \cellcolor[HTML]{165128} 9.7m & \cellcolor[HTML]{8ac794} 82.3m & \cellcolor[HTML]{FF9999} T.O & \cellcolor[HTML]{60b472} 37.1m & \cellcolor[HTML]{FF9999} T.O & \cellcolor[HTML]{FF9999} T.O \\
TIF006 & \cellcolor[HTML]{165128} 1.9m & \cellcolor[HTML]{8ac794} 82.3m & \cellcolor[HTML]{b2dab7} 15.5h & \cellcolor[HTML]{60b472} 44.1m & \cellcolor[HTML]{ffffff} 22.3h & \cellcolor[HTML]{d8edda} 20.6h \\
TIF007 & \cellcolor[HTML]{ffffff} 6.8m & \cellcolor[HTML]{60b472} 18s & \cellcolor[HTML]{8ac794} 21s & \cellcolor[HTML]{165128} 8s & \cellcolor[HTML]{b2dab7} 25s & \cellcolor[HTML]{d8edda} 29s \\
TIF008 & \cellcolor[HTML]{165128} 7.6m & \cellcolor[HTML]{FF9999} T.O & \cellcolor[HTML]{60b472} 16h & \cellcolor[HTML]{8ac794} 16.6h & \cellcolor[HTML]{b2dab7} 20h & \cellcolor[HTML]{FF9999} T.O \\
TIF009 & \cellcolor[HTML]{165128} 20.2m & \cellcolor[HTML]{FF9999} T.O & \cellcolor[HTML]{8ac794} 4.7h & \cellcolor[HTML]{b2dab7} 6.2h & \cellcolor[HTML]{d8edda} 16.9h & \cellcolor[HTML]{60b472} 2.3h \\
TIF012 & \cellcolor[HTML]{8ac794} 19.6m & \cellcolor[HTML]{d8edda} 90.8m & \cellcolor[HTML]{60b472} 6m & \cellcolor[HTML]{165128} 2.6m & \cellcolor[HTML]{ffffff} 1.7h & \cellcolor[HTML]{b2dab7} 53m \\
TIF014 & \cellcolor[HTML]{60b472} 12.8m & \cellcolor[HTML]{d8edda} 87.2m & \cellcolor[HTML]{8ac794} 13.2m & \cellcolor[HTML]{165128} 10.3m & \cellcolor[HTML]{b2dab7} 53.9m & \cellcolor[HTML]{ffffff} 2.3h \\
XML001 & \cellcolor[HTML]{165128} 3.4m & \cellcolor[HTML]{FF9999} T.O & \cellcolor[HTML]{60b472} 14.8h & \cellcolor[HTML]{8ac794} 20.9h & \cellcolor[HTML]{d8edda} 22.1h & \cellcolor[HTML]{b2dab7} 21.4h \\
XML002 & \cellcolor[HTML]{165128} 1.8m & \cellcolor[HTML]{FF9999} T.O & \cellcolor[HTML]{60b472} 17.7h & \cellcolor[HTML]{8ac794} 22.9h & \cellcolor[HTML]{FF9999} T.O & \cellcolor[HTML]{FF9999} T.O \\
XML003 & \cellcolor[HTML]{165128} 2.8m & \cellcolor[HTML]{FF9999} T.O & \cellcolor[HTML]{60b472} 10.6h & \cellcolor[HTML]{d8edda} 17.9h & \cellcolor[HTML]{b2dab7} 16.1h & \cellcolor[HTML]{8ac794} 14.3h \\
XML009 & \cellcolor[HTML]{165128} 7.9m & \cellcolor[HTML]{ffffff} 5.3h & \cellcolor[HTML]{d8edda} 78m & \cellcolor[HTML]{b2dab7} 72.6m & \cellcolor[HTML]{60b472} 11.4m & \cellcolor[HTML]{8ac794} 19.6m \\
XML010 & \cellcolor[HTML]{165128} 8.5m & \cellcolor[HTML]{FF9999} T.O & \cellcolor[HTML]{FF9999} T.O & \cellcolor[HTML]{FF9999} T.O & \cellcolor[HTML]{FF9999} T.O & \cellcolor[HTML]{FF9999} T.O \\
XML012 & \cellcolor[HTML]{FF9999} T.O* & \cellcolor[HTML]{FF9999} T.O & \cellcolor[HTML]{8ac794} 23.1h & \cellcolor[HTML]{FF9999} T.O & \cellcolor[HTML]{60b472} 19.8h & \cellcolor[HTML]{165128} 6.5h \\
XML017 & \cellcolor[HTML]{ffffff} 3.4m & \cellcolor[HTML]{8ac794} 13s & \cellcolor[HTML]{b2dab7} 13s & \cellcolor[HTML]{d8edda} 61s & \cellcolor[HTML]{165128} 6s & \cellcolor[HTML]{60b472} 8s \\
\bottomrule
\end{tabular}
}
\end{table}

\subsubsection{Time-to-Reach and Reproducibility}
Due to space constraints, detailed Time-to-Reach (TTR) and reproducibility analyses
are provided in \autoref{sec:app-ttr} and \autoref{sec:app-consistency}.

\begin{table}[tbp]
\centering
\small
\caption{API cost from Cursor per vulnerability}
\label{tab:cost}
\begin{tabular}{l r r r r r}
\toprule
\textbf{Project} & 
  \textbf{Input} & 
  \textbf{Cache} & 
  \textbf{Output} & 
  \textbf{Total} & 
  \textbf{Cost} \\
& 
  \textbf{Tokens} & 
  \textbf{Read} & 
  \textbf{Tokens} & 
  \textbf{Tokens} & 
  \textbf{(\$)} \\
\midrule
libpng     & 64.0k     & 2737k & 20k & 2875k & 1.53 \\
openssl    & 49.044k   & 2109k & 11k & 2170k & 0.99 \\
php        & 160.120k  & 7261k & 32k & 7455k & 3.28 \\
libsndfile & 95.074k   & 3754k & 21k & 3870k & 1.80 \\
sqlite3    & 154.091k  & 4730k & 29k & 4915k & 2.45 \\
poppler    & 48.035k   & 1923k & 18k & 1989k & 1.04 \\
libxml2    & 164.0k    & 98k   & 27k & 5214k & 2.66 \\
lua        & 76.054k   & 1474k & 24k & 1574k & 1.09 \\
libtiff    & 182.089k  & 6704k & 57k & 6943k & 3.55 \\
\bottomrule
\end{tabular}
\end{table}

\subsubsection{Cost Analysis}
\autoref{tab:cost} summarizes the API cost incurred by \sys
when generating PoVs for each vulnerability across all projects.
This expenditure correlates with tool invocation frequencies (\autoref{fig:tool-calling-stats}),
confirming that extensive analysis drives token consumption.
On average, \sys consumes 2.18 million tokens and incurs a cost of \$1.83 per vulnerability.
Given the time-sensitive nature of vulnerability verification,
this expenditure remains acceptable compared to
the substantial computational resources required by 24-hour greybox fuzzing campaigns.
Furthermore, \sys demonstrates robust scalability:
\cc{PHP} (1192K LOC) incurs only 3$\times$ the cost of \cc{lua} (23K LOC),
showing that agentic reasoning selectively retrieves relevant context
rather than exhaustively processing the entire massive codebase.

\begin{findingbox}
\textbf{A1:}
\sys significantly outperforms SOTA PoV generation approaches.
It triggered more vulnerabilities, more unique vulnerabilities,
faster, and more consistently than all baselines.
\end{findingbox}

\subsection{RQ2. Strengths and Limitations}
\label{sec:eval:strengths-limitations}

We analyze the 17 vulnerabilities only \sys triggers,
and 12 vulnerabilities it fails to trigger but AFL++ triggers,
to perform in-depth validation of our hypotheses and
to identify unique capabilities and unaddressed challenges.
Overall, our analysis reveals four key strengths:
(S1) autonomous reachability analysis (6 CVEs),
(S2) property-based testing for structural invariant preservation (3 CVEs),
(S3) hierarchical semantic constraint solving (7 CVEs), and
(S4) semantic constraint solving for state machines (2 CVEs).
We also identify three limitations stemming from tensions between
LLM training objectives and vulnerability exploitation requirements:
(L1) systemic scope limitation where agents miss distant code dependencies (2 CVEs),
(L2) constraint search gap in mapping semantic constraints to parameter spaces (3 CVEs), and
(L3) construct validity bias producing exclusively well-formed inputs (4 CVEs).
The remaining three CVEs failed to trigger
due to agent execution errors rather than methodological limitations:
\cc{SND006} and \cc{SND007} timed out during environment setup,
while \cc{SND024} failed due to a generator implementation bug
compounded by a diagnostic misattribution that exhausted the budget.
Detailed case studies are discussed in \autoref{sec:app-strengths}
and \autoref{sec:app-limitations}.

\PP{Strength 1: Autonomous Reachability Analysis}
Six vulnerabilities (\cc{PHP001}, \cc{PHP003}, \cc{PHP010}, \cc{PHP013}, \cc{PHP014}, \cc{SSL006})
cannot be reached through Magma-provided harnesses.
\sys identified these reachability barriers via autonomous call graph analysis,
and switched to the correct harnesses that are already included in the codebase (for PHP vulnerabilities),
or synthesized custom harnesses (for SSL006).
This demonstrates \sys's unique capability:
autonomously identifying and fixing harness inadequacies that remained hidden to prior work.
This capability proves critical for real-world scenarios where
newly introduced features lack corresponding fuzz harnesses
\cite{ispoglou2020fuzzgen,babic2019fudge,zhang2024effective, yang2025harnessagent}.
This also validates H1: \sys can infer reachability barriers and overcome them autonomously.

\PP{Strength 2: Format Invariants Preservation}
Three vulnerabilities (\cc{PDF005}, \cc{PDF022}, \cc{SND016}) require maintaining
input structural invariants across nested layers,
where single-field mutations will corrupt global validity.
Baseline fuzzers failed because they cannot maintain these invariants.
\sys synthesizes parameterized generators encoding structural dependencies as compositional functions,
enabling atomic multi-field updates while preserving format validity.
These cases validate H2: property-based testing systematically searches
solution spaces without breaking structural invariants.

\PP{Strength 3: Hierarchical Semantic Constraints Solving}
Seven vulnerabilities (\cc{LUA001}, \cc{PDF002}, \cc{PDF008}, \cc{PDF009}, \cc{PDF012}, \cc{SQL010}, \cc{XML010}) require
solving nested constraints coupling syntactic structures with semantic values and runtime state.
These are reachable by fuzzers, but solving hierarchical constraints exceeds random mutation capabilities.
\sys resolves constraint hierarchies via hypothesis-driven exploration in PLAN
and structured synthesis in IMPLEMENT,
supporting the second half of H1: \sys can successfully map semantic-level
constraints to input parameter spaces.

\PP{Strength 4: Semantic Constraint Solving}
Two vulnerabilities (\cc{SQL007}, \cc{SND016}) require driving execution through
program state machines (schema reload, parser selection).
A PoV generator must (1) infer state machines,
(2) reason about transitions to reach vulnerable states, and
(3) generate inputs driving these transitions.
\sys successfully overcame these via autonomous backward dependency analysis,
LLM-based reasoning for semantic constraints,
and synthesized parameterized generators driving state transitions.
This validates H1: \sys infers and solves reachability barriers distance metrics cannot capture.

\PP{Limitation 1: Systemic Scope Limitation}
During PLAN, \sys focuses on local vulnerability contexts but misses distant dependencies.
Two vulnerabilities (\cc{XML012}, \cc{PDF011})
are triggered by side effects from components dismissed as irrelevant.
Agents prune ``irrelevant'' code paths to manage cognitive load,
missing critical indirect interactions (character encoding, annotation processing).

\PP{Limitation 2: Constraint Search Gap}
During IMPLEMENT, \sys identifies high-level semantic constraints
but fails to map them to specific parameter spaces.
Three vulnerabilities (\cc{SSL009}, \cc{SSL001}, \cc{PDF006})
depend on logical conditions satisfiable by multiple byte-level encodings.
\sys explored standard encodings but failed to enumerate
alternative ASN.1 forms, unusual tag values (e.g., 266 for \cc{V\_ASN1\_NEG\_ENUMERATED}),
or integer boundaries (\cc{\textpm 32768}).
AFL++ discovered alternatives via seed corpus examination.
The agent lacked explicit enumeration heuristics or domain-specific knowledge
for systematically exploring complex format parameter spaces.

\PP{Limitation 3: Construct Validity Bias}
During EXECUTE, \sys-synthesized generators produce exclusively well-formed inputs
despite workflow rules instructing malformed structure prioritization.
Four vulnerabilities (\cc{TIF001}, \cc{TIF002}, \cc{SQL002}, \cc{SQL014})
require violating structural invariants.
LLM training objectives align models to produce syntactically valid outputs.

\begin{findingbox}
\textbf{A2:}
\sys's strengths stem from LLM-driven semantic constraint inference,
constraint-to-parameter-space encoding, and property-based testing,
supporting the validation of H1 and H2.
Its limitations stem from pruning distant dependencies,
incomplete parameter-space enumeration,
and generator bias toward valid constructs.
\end{findingbox}

\subsection{RQ3. Ablation Study}
\label{sec:eval:ablation}

\subsubsection{Impacts of System Components and Design Choices}
\sys extends an off-the-shelf LLM agent (cursor-cli) with PoV-specific workflow,
additional MCP tools, persistent memory, and property-based testing.
To isolate contributions from different components and design choices,
we evaluate four progressively enhanced variants
measuring impacts of agentic design, program analysis tools, and feedback granularity:

\begin{itemize}[leftmargin=*]
  \item {\bf LLM only}: Fixed prompt template
  with one-shot PoV generation to evaluate baseline LLM capability with
  fixed prompt context extracted from static analysis
  and no dynamic tool calling for information gathering.
  
  \item {\bf \cursor}: Cursor agent with built-in tools
  (\cc{Grep}, \cc{Glob}, \cc{Shell}, \cc{Read}).
  Measures agentic iteration baseline contribution beyond one-shot prompting.
  The agent is launched using the system prompt in \autoref{lst:cursor-prompt}.
  
  \item {\bf \cursorFB}: Extends \cursor by incorporating execution feedback
  indicating target reachability and vulnerability triggering status.
  Tests coarse-grained feedback value for hypothesis refinement.
  
  \item {\bf \cursorTools}: Extends \cursorFB by integrating directed-fuzzing-specific MCP tools from \sys
  (\cc{get\_callers}, \cc{get\_callees}, \cc{get\_reaching\_routes}, \cc{detect\_deviation}).
  Isolates static analysis contribution to constraint collection.

  \item {\bf Full \sys}: Extends \cursorTools by integrating workflow, persistent memory,
  and property-based testing (\cc{fuzz} tool).
\end{itemize}

All variants employ Claude-Sonnet 4.5 as the foundational LLM backbone.
We execute LLM-only across five runs, merging findings to reduce variance,
capturing the upper bound of one-shot prompting capability.
All agentic variants (\cursor, \cursorFB, \cursorTools, and \sys) execute as single runs.
The cursor agent utilizes a specialized system prompt
that instructs the agent to analyze program structure, locate bug predicates, and generate PoV inputs.

\begin{figure}[tbp]
\centering
\includegraphics[width=\columnwidth]{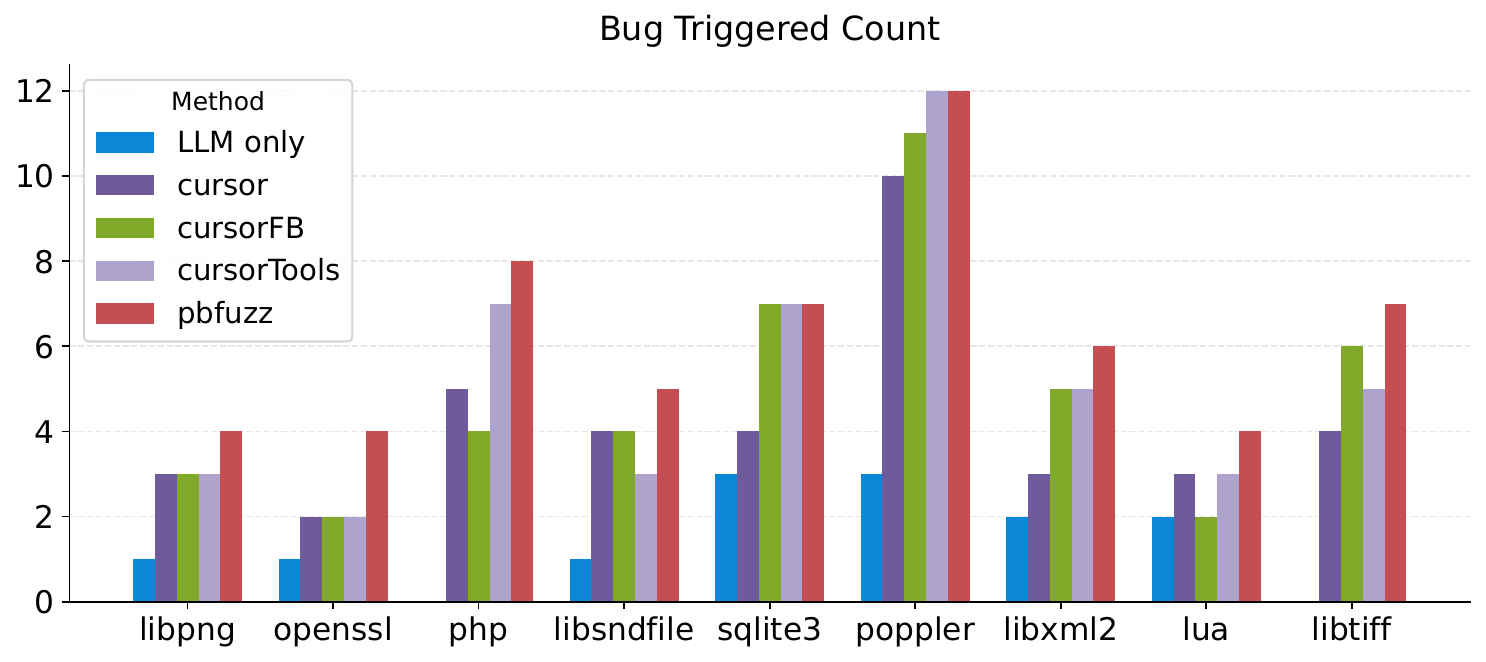}
\caption{
CVE coverage across different system variants.
Each bar group represents one project from Magma.
}
\label{fig:ablation-study}
\vspace{-1em}
\end{figure}

\autoref{fig:ablation-study} shows progressive improvements
on the number of vulnerabilities triggered,
from 13 vulnerabilities (pure LLM) to 57 vulnerabilities (full \sys).
We explain each performance jump
by analyzing root causes of baseline failures.

$\bullet$
\cursor (38 vulnerabilities) outperformed pure LLM (13 vulnerabilities) by 192.3\%.
LLM failures stem mostly from static context limitations:
fixed prompts lack PUT-specific dependency chains
and type information required for valid reachability constraints.
In contrast, \cursor leverages autonomous tool orchestration---
agents invoke program analysis tools to retrieve
constraint-relevant information on-demand,
enabling formulation of valid input constraints.

$\bullet$
\cursorFB (44 vulnerabilities) improved over \cursor by 15.8\%.
The open-loop agent hallucinated successful generation based on
the LLM's own plausibility reasoning without verification.
Even coarse-grained feedback---binary signals for reachability and triggering---
can prevent the agent from converging on invalid hypotheses.
Hence, the agent can continue to refine its hypotheses until successfully
triggering the vulnerability.

$\bullet$
\cursorTools (47 vulnerabilities) improves 6.8\% over \cursorFB.
We hypothesize that all MCP tools contribute.
Upon subsequent analysis (\autoref{sec:tool-calling-stats}),
we identified specific tools driving this improvement.
First, for control-flow reachability analysis, we found that the invocation of
\cc{get\_reaching\_routes} with Magma-supplied initial seeds
enables the agent to quickly formulate accurate reachability hypotheses.
Second, by combining interactive GDB-based runtime state inspection and
\cc{detect\_deviation}, the agent can quickly pinpoint the mistakes in
its previous hypotheses and refine them effectively.

$\bullet$
Full \sys (57 vulnerabilities) achieves 21.3\% improvement.
This improvement stems from two main sources.
The workflow and persistent memory components help the agent decompose tasks,
ensure the agent stays on track during multi-phase reasoning without deviation,
and retain validated knowledge across iterations.
Second, the property-based-testing (PBT)-based fuzzer overcomes the
non-negligible latency inherent to direct LLM-based generation.
By systematically sampling typed parameter spaces derived from LLM-inferred
constraints with \emph{high-throughput},
the agent can solve the constraints much faster.
This architectural separation---agents reason, PBT solves---
fundamentally bridges semantic reasoning and PoV generation.

\subsubsection{Performance of PIER Stages}
\label{sec:eval:ablation:stage-level}

In this subsection, we analyze the performance of individual
PLAN--IMPLEMENT--EXECUTE--REFLECT (PIER) stages (\autoref{sec:overview}).
In particular, we measure, per round,
how accurately the agent infers bug-critical constraints (PLAN),
how completely those constraints enter the typed parameter space (IMPLEMENT),
and whether the resulting generator reaches and triggers the vulnerability (EXECUTE).

\PP{Scope}
We restrict this analysis to vulnerabilities
requiring at least three PIER rounds to trigger
({\cc{PDF005}, \cc{PDF012}, \cc{PDF019},
\cc{PHP001}, \cc{PHP004}, \cc{PHP009}, \cc{PHP010}, \cc{PHP011},
\cc{SSL003}, \cc{SSL020}, \cc{TIF014}}).
Targets that converge in one round expose no stage-level evolution;
and untriggered targets are reserved for separate failure analysis.

\PP{Ground-Truth Protocol}
For each vulnerability, we manually extract
a minimal set of \emph{critical ground-truth constraints}---
conditions whose violation prevents the PoV
from reaching or triggering the bug.
Extraction follows three principles:
\emph{necessity}, %
\emph{minimality}, %
and \emph{semantic abstraction}
(e.g., ``declared byte count exceeds actual data length''
rather than a specific raw byte offset).
Against this ground truth,
we score each PIER round with stage-specific metrics.

\begin{itemize}[leftmargin=1.5em]
\item \emph{PLAN} stage generates a set of reaching and triggering constraints,
so we measure two metrics:
\textbf{precision} (the fraction of inferred constraints that are correct) and
\textbf{recall} (the fraction of ground-truth constraints that are covered).

\item \emph{IMPLEMENT} stage encodes inferred constraints into a typed parameter space,
so we measure two metrics:
\textbf{coverage} (the fraction of inferred constraints that are parameterized) and
\textbf{fidelity} (the fraction of parameterized constraints that are correctly encoded).

\item \emph{EXECUTE} stage runs the generated PoV and observes two binary outcomes:
\textbf{reachability} (whether the input reached the target code location) and
\textbf{triggering} (whether the input triggered the vulnerability).
\end{itemize}

\begin{figure}[t]
\centering
\includegraphics[width=\columnwidth]{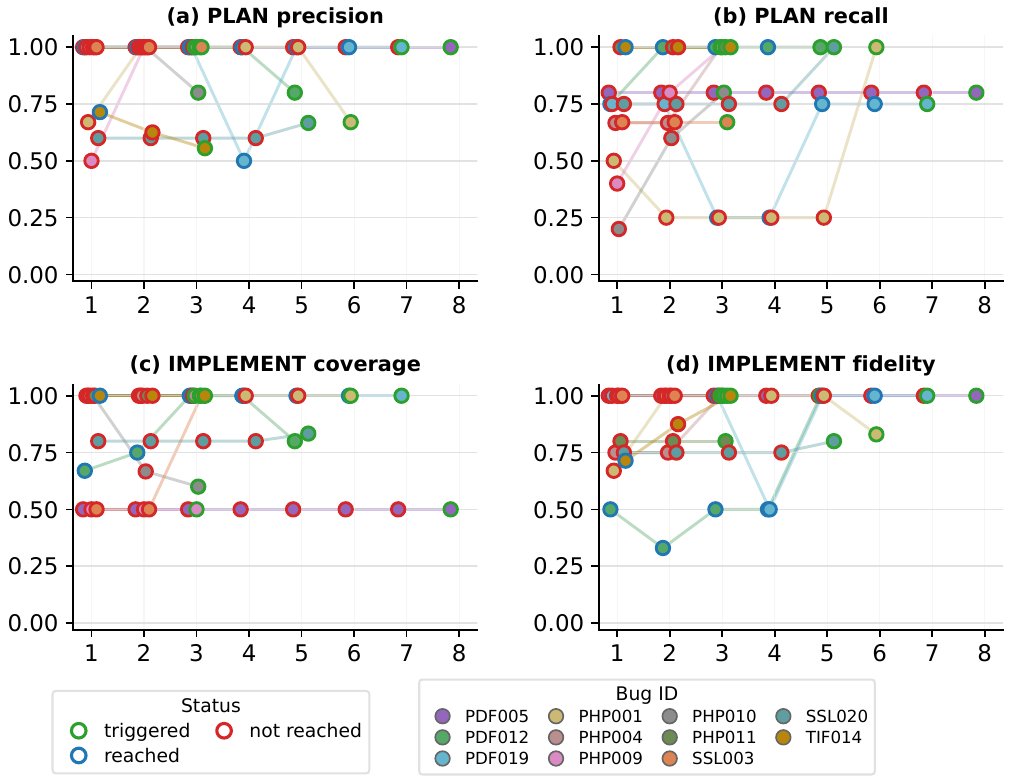}
\caption{Stage-level PIER reliability across $11$ targets
that require at least three PIER rounds to trigger.
The $x$-axis is the PIER round number;
the marker fill color and trajectory line color encode bug identity;
the marker ring color encodes EXECUTE outcome.
Each connected marker chain is one target over rounds.
A PIER trajectory traces from red/blue rings
in early rounds toward green rings in the final round.}
\label{fig:stage-level-ablation}
\vspace{-1em}
\end{figure}

\PP{Aggregated Results}
\autoref{fig:stage-level-ablation} plots all key metrics
across the 11 targets and their 49 PIER rounds,
where each chain of markers traces one target's trajectory across rounds.
Across all four panels, the general pattern is final-round improvement:
while individual trajectories can be non-monotone,
marker ring (edge) colors generally shift from red (not reached) and
blue (reached but not triggered) in early rounds toward green (triggered)
in final rounds.
Triggering successes rise from $0/11$ in round~1 to $11/11$ in the final round;
$\mathit{plan\_recall} \geq 0.8$ rises from $3/11$ to $9/11$;
and mean $\mathit{implement\_fidelity}$ improves
from $0.835$ to $0.948$ (median $1.0$).
These results indicate that final-round success depends on iterative repair
rather than one-shot constraint inference.
This aggregate repair pattern supports \textbf{OC3}
(Fine-Grained Execution Feedback,~\autoref{sec:intro}):
without per-round failure signals fed back through REFLECT,
the agent has no basis for recovering missing constraints.
Conditionally, among the $9$ final rounds
with actionable PLAN recall ($\mathit{plan\_recall} \geq 0.8$),
all $9$ synthesize a triggering generator;
restricting further to high IMPLEMENT quality
($\mathit{implement\_coverage} \geq 0.8$
and $\mathit{implement\_fidelity} \geq 0.8$),
all $6$ qualifying rounds also trigger.

\PP{Individual Stage Analysis}

$\bullet$~\emph{PLAN Stage.}
PLAN-stage precision is high from round~1
(panel~(a), median $1.0$ and the majority of targets remain near $y{=}1.0$),
whereas recall starts low and climbs gradually
(panel~(b), first-round mean $0.681$, final-round mean $0.911$).
This pattern indicates that \sys uses a conservative inference strategy:
it only proposes constraints when it has explicit evidence from code inspection.
Consequently, early rounds yield few inferred constraints,
but most of them are correct, driving precision high.
As PIER rounds progress, the agent proposes additional missing constraints.
This pattern matches \textbf{L1} Systemic Scope Limitation
(\autoref{sec:eval:strengths-limitations}):
agents prune long call chains to manage cognitive load,
and the PIER loop compensates by iteratively recovering
the missing conditions.

$\bullet$~\emph{IMPLEMENT Stage.}
Fidelity of the IMPLEMENT stage has the highest overall performance
(panel~(d), $\geq 0.8$ on $11/11$ final rounds, $8/11$ perfect, median $1.0$).
This result confirms the observation in~\cite{zhang2025low}:
LLMs encode common input-format conventions sufficiently precisely
to translate semantic constraints into parameter spaces.
This evidence supports our design choice of hierarchical separation
between LLM-driven semantic reasoning and PBT-based search:
if constraints are encoded in the typed parameter space correctly,
PBT can solve them efficiently and faithfully.
However, IMPLEMENT coverage (panel~(c)) remains highly variable
and its trajectory can \emph{decline} as PLAN recall improves,
showing newly inferred constraints were not always parameterized.
Our in-depth analysis \emph{confirms} \textbf{L2} Constraint Search Gap
(\autoref{sec:eval:strengths-limitations}):
beyond the missing alternative encodings identified in RQ2,
stage-level data reveals a distinct second manifestation:
when the agent discovers structural reachability conditions late,
it often hardcodes them directly into the generator code
rather than exposing them as search dimensions for PBT.
Hardcoded conditions become rigid all-or-nothing gates:
the generator either satisfies them exactly or fails completely,
with no intermediate search space to traverse.

\begin{figure}[t]
\centering
\includegraphics[width=\columnwidth]{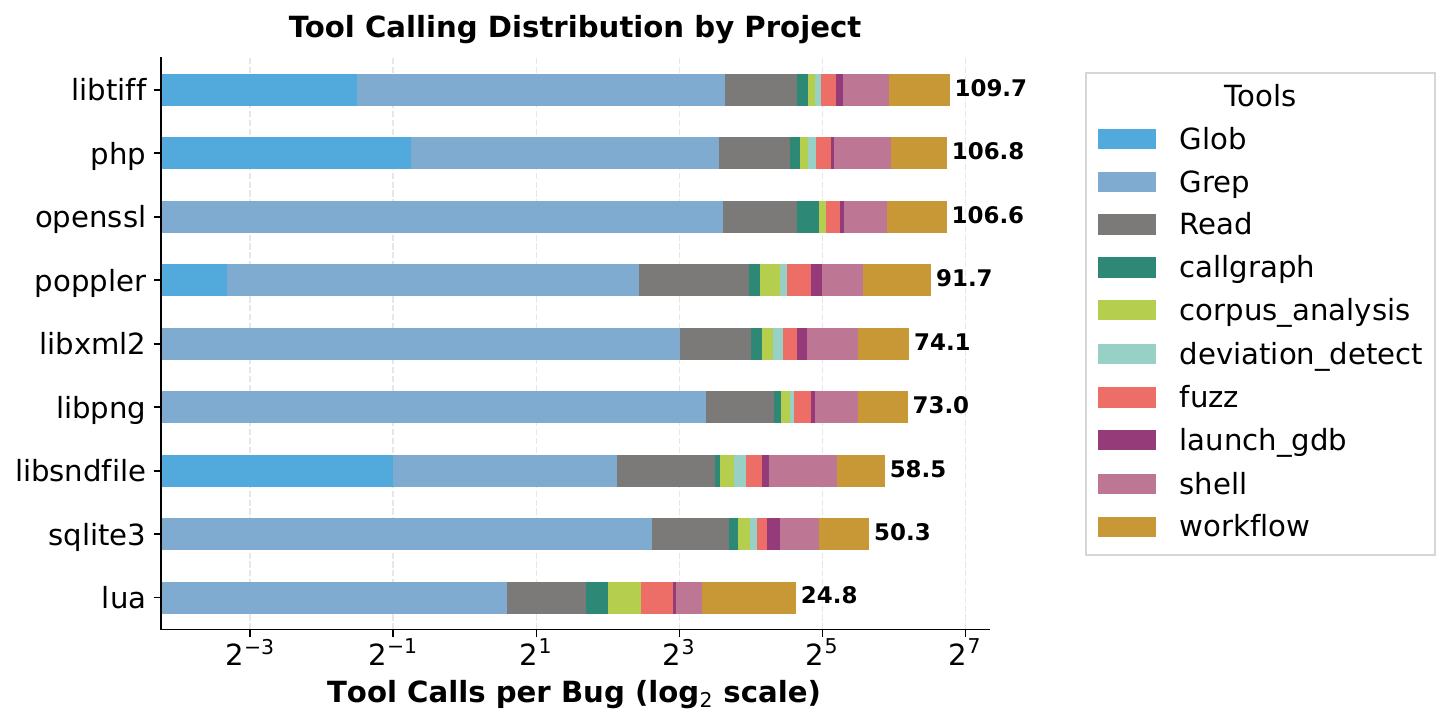}
\caption{
Average tool invocations per vulnerability, by project.
Within each bar, segments represent tool proportion.
X-axis uses log$_2$ scale.
}
\label{fig:tool-calling-stats}
\vspace{-1em}
\end{figure}

\subsubsection{Tool Calling Frequencies}
\label{sec:tool-calling-stats}

\autoref{fig:tool-calling-stats} quantifies tool invocation patterns across projects.
Codebase exploration tools (Glob, Grep, Read) account for
21.0\% (2,782/13,250) of invocations,
confirming that constraint inference fundamentally requires source code understanding.
Generally, larger and more complex codebases demand proportionally higher analysis effort---
\cc{libtiff} (109.7 calls/bug), \cc{PHP} (106.8), and \cc{OpenSSL} (106.6)
required 4.3--4.4$\times$ more invocations than \cc{lua} (24.8).
Critically, while \cc{PHP}'s codebase size exceeds \cc{lua} by 51.8$\times$ (1192K vs. 23K LOC),
the tool invocations scale sublinearly (4.3$\times$).
This sublinear growth demonstrates robust scalability:
\sys's reasoning ability allows it to focus on relevant code regions,
avoiding exhaustive exploration of the entire codebase.

Workflow management constitutes 43.8\% of calls.
Closer examination reveals this primarily results from failed attempts
where the agent performed more iterations and more frequent phase changes
to refine its hypotheses.
Notably, deviation detection represents merely 1.7\% (222 invocations).
This minimal deviation analysis aligns with Magma benchmark characteristics:
target sites are typically reachable through straightforward input formats,
reducing the necessity for reachability constraint refinement.

\subsubsection{Model Variant Analysis}
We evaluated the architecture's performance with different LLM backbones
to isolate architectural contributions from model capabilities.
Detailed comparison across Claude Sonnet-4.5, Claude Sonnet-4.5-Thinking, GPT-5, and Grok-4
is provided in \autoref{sec:app-model-variants}.
Our analysis demonstrates that superior code reasoning capabilities
and tool-use proficiency enable effective constraint inference and solver synthesis.
Moreover, structured workflow design outperforms model-level reasoning
for complex multi-phase tasks.

\begin{findingbox}
\textbf{A3:}
Simple adoption of LLMs or LLM-based agents is insufficient.
\sys's key innovations---its workflow, MCP tools, persistent memory,
and the adoption of property-based testing---all contribute to its superior performance.
Stage-level diagnostics further attribute these gains
to two complementary properties:
high-precision constraint inference enabled by MCP-tool-backed source retrieval,
and high-fidelity generator synthesis enabled by the typed parameter space;
the residual bottleneck lies in expanding parameter-space coverage
of structural reachability conditions.
\end{findingbox}

\subsection{Real-world Application}
\label{sec:eval:real-world-application}

We further evaluate \sys as a practical PoV generation workflow
for real-world vulnerabilities outside benchmark replay.

\subsubsection{1-Day Vulnerability Reproduction}
\label{sec:eval:one-day-reproduction}

One-day vulnerability reproduction tests whether \sys can generate PoVs
from public CVE semantics and patch information alone.
We selected the \cc{FFmpeg} project as the target because it is extensively fuzzed
by OSS-Fuzz~\cite{ossfuzz} and is widely adopted in prior fuzzing evaluations~\cite{li2021unifuzz,fu2023autofz,zhang2025low}.
This setting evaluates \sys on a mature target with extensive prior fuzzing coverage.

\PP{Target}
For this study, we extended the driver with an INIT agent.
The INIT agent identifies the vulnerable revision, builds \cc{FFmpeg},
selects sanitizers, writes the execution command,
derives patch-based reaching and triggering oracles,
and retries setup when verification fails.
This extension completes the upstream environment setup required
by real-world CVE reproduction.
We treat it as workflow engineering that operationalizes
the prerequisites stated in~\autoref{sec:overview},
leaving the inner PIER loop responsible for PoV generation.

\PP{Selection Protocol}
To ensure the objectivity and validity of this evaluation,
we audited all 2026 CVEs mentioning \cc{FFmpeg} at the time of writing
and applied the following filtering criteria.
First, we conducted an extensive search to
\emph{exclude CVEs with public ready-to-run PoVs}.
This criterion reduces the risk that the reproduction task
is solvable through model memorization.
Second, we retained only upstream \cc{FFmpeg} vulnerabilities that are reachable
through file or stream inputs,
controlled by structured media fields,
and supported by component-level semantic details.
\autoref{sec:app-ffmpeg-selection} summarizes the resulting decision table,
and three qualified CVEs are shown in~\autoref{tab:ffmpeg-reproduction-cves}.

\begin{table}[t]
\centering
\scriptsize
\caption{Qualified FFmpeg CVEs for 1-day reproduction.}
\label{tab:ffmpeg-reproduction-cves}
\begin{tabular}{@{}lp{0.8cm}p{2.15cm}p{1.3cm}p{1.45cm}@{}}
\toprule
\textbf{CVE} &
\textbf{Date} &
\textbf{Root Cause} &
\textbf{Scope} &
\textbf{Patch / PoV} \\
\midrule
\cc{CVE-2026-40962} &
2026-04-15 &
CENC subsample integer overflow in \cc{libavformat/mov.c} &
FFmpeg before 8.1; MP4/MOV &
Patch available; no public PoV. \\
\cc{CVE-2026-6385} &
2026-04-15 &
DVD subtitle fragment reassembly signed overflow &
MPEG-PS/VOB input &
Upstream fix unresolved; no public PoV. \\
\cc{CVE-2026-30997} &
2026-04-13 &
AV1 \cc{read\_global\_param()} OOB read when \cc{primary\_ref\_frame == 7} &
FFmpeg up to 8.0.1; AV1 IVF &
Patch available; no public PoV. \\
\bottomrule
\end{tabular}
\vspace{-1em}
\end{table}

\PP{Results and Analysis}
\sys reproduced all three CVEs with sanitizer-confirmed PoVs.
We analyze the reproduction trajectories below.

$\bullet$ CVE-2026-40962: early generators failed to reach
the bug location (the CENC subsample loop) that checks encrypted MP4 segment sizes,
due to incomplete MP4 encryption metadata or an incorrect binary path.
After PIER round~10, the oracle reported \emph{reached} but not
\emph{triggered}.
The agent then revised the generator toward the \cc{cbc1}
encryption mode and boundary-sized protected data;
PIER round~17 reached and triggered the oracle in 7/10 tests,
and ASan confirmed the resulting AES access past a 64-byte buffer.

$\bullet$ CVE-2026-6385: initial attempts terminated before processing
the intended $2$GB DVD subtitle stream,
which indicated an MPEG-PS reachability failure.
REFLECT identified a missing PES header-data-length byte,
the field that specifies where the subtitle payload begins:
adding the final \cc{0x00} after \cc{0x80 0x00}
delivered the DVD subtitle payload to \cc{dvdsub\_parse()}.
The final generator then emitted 32,770 subtitle fragments
with \cc{packet\_len=2147483583},
raising \cc{packet\_index} to 2,147,450,870;
adding 65,531 then overflowed signed \cc{int},
which bypassed the fragment bounds check.
UBSan reported the same arithmetic violation.

$\bullet$ CVE-2026-30997: %
across 16 PIER rounds, \sys's persistent memory (\cc{workflow\_state.md})
converged on three semantic preconditions:
an AV1 INTER frame, \cc{error\_resilient\_mode=1},
and global motion for a reference frame.
These conditions constrain \cc{primary\_ref\_frame} to 7
and route decoding into \cc{read\_global\_param()},
the vulnerable function that indexes \cc{ref\_frame\_idx}.
By round~13, the agent had constructed a valid AV1 IVF bitstream
satisfying these conditions.
UBSan then reported the \cc{ref\_frame\_idx[7]} OOB read.

\subsubsection{New Bug Discovery}
\label{sec:eval:new-bug-discovery}

We additionally evaluate Bento4,
a widely used MP4 library from prior fuzzing studies~\cite{li2021unifuzz,lin2024deepgo,xiang2024critical},
to assess whether \sys can discover previously unknown vulnerabilities.

To assess vulnerability discovery on an additional real-world target,
we applied \sys to Bento4~\cite{bento4},
a widely-deployed MP4 processing library with 218K lines of code.
Following \sys's prerequisites (\autoref{sec:overview}),
we first compiled the project and analyzed the binary to identify parser APIs.
We then employed the Cursor agent to perform static analysis
on suspicious APIs handling untrusted input,
hypothesizing potential bug triggering conditions
(e.g., missing bounds checks, unconstrained loops).
For each hypothesized vulnerability,
we instrumented the code with reaching and triggering signals,
then executed \sys to generate PoV inputs.
\sys confirmed three previously unknown vulnerabilities.
\autoref{tab:new-bugs} summarizes these findings.

\begin{table}[tbp]
\centering
\scriptsize
\caption{New vulnerabilities discovered in Bento4}
\label{tab:new-bugs}
\begin{tabular}{@{}cp{1.2cm}p{5.8cm}@{}}
\toprule
\textbf{\#} & \textbf{Type} & \textbf{Location \& Root Cause} \\
\midrule
BUG1 & Global OOB & Ap4Ac3Parser.cpp:46 - Array[3][38] indexed by [0-3][0-63] \\
BUG2 & Heap OOB & Ap4Utils.cpp:447 - ReadGolomb without buffer check \\
BUG3 & DoS & Ap4SttsAtom.cpp:81 - entry\_count loop without validation \\
\bottomrule
\end{tabular}
\vspace{-1em}
\end{table}

\textbf{Bug 1}: When AC3 parser accesses
frame size array without bounds check.
The array is [3][38] but indices may be
2-bit (0-3) and 6-bit (0-63) values.
\sys succeeded because:
(1) PLAN phase identified the dimension
mismatch via source reading,
extracting the bug predicate
\cc{m\_Fscod$\geq$3 OR m\_Frmsizecod$\geq$38};
(2) IMPLEMENT phase encoded this as typed parameter space
with \cc{fscod} and \cc{frmsizecod} ranges;
(3) EXECUTE phase generated PoV with \cc{fscod=3}
(AC3 RESERVED value) in first iteration,
triggering global buffer overflow in 2.1 minutes.

\textbf{Bug 2} occurs when \cc{ReadGolomb()} at Ap4HevcParser.cpp:149
loops \cc{while(!bits.ReadBit())} during exp-golomb decoding
without checking buffer bounds,
causing \cc{ReadCache()} at Ap4Utils.cpp:447 to access beyond heap memory.
\sys succeeded because:
(1) PLAN phase traced backward from \cc{ReadCache()}
to identify missing bounds check in \cc{ReadGolomb()};
(2) IMPLEMENT phase encoded constraints as parameter space
with NAL structure (start code, type=32 for VPS) and malformed payload patterns;
(3) EXECUTE phase generated 32-byte truncated VPS NAL
with \cc{0xFF} payload exhausting buffer during profile parsing,
triggering heap overflow in 3.2 minutes.

\textbf{Bug 3} occurs at Ap4SttsAtom.cpp:81
where \cc{stream.ReadUI32} followed by
\cc{while(entry\_count--)} loops without validation,
causing 2.1 billion iterations when \cc{entry\_count=0x7FFFFFFF}.
\sys succeeded because:
(1) PLAN phase identified loop pattern without reasonableness check;
(2) IMPLEMENT phase encoded MP4 box hierarchy constraints
(ftyp/moov/trak/mdia/minf/stbl structure) as parameter space
to preserve format validity;
(3) EXECUTE phase generated minimal valid MP4
with stts atom containing \cc{entry\_count=0x7FFFFFFF} but zero actual entries,
triggering timeout in 1.5 minutes.

These vulnerabilities validate strengths identified in RQ2.
\textbf{Bug 1} demonstrates \textbf{S3 (hierarchical semantic constraint solving)}:
inferring array dimension mismatches (\cc{[3][38]} vs \cc{[0-3][0-63]})
and targeting boundary values (\cc{fscod=3}) that coverage feedback cannot distinguish.
\textbf{Bugs 2-3} demonstrate \textbf{S2 (format invariants preservation)}:
\sys synthesized generators maintaining valid container structures
(HEVC NAL headers, MP4 box hierarchies)
while targeting vulnerabilities through malformed payloads or boundary values,
whereas random mutation corrupts structural validity.
\sys's constraint-driven generation systematically reasons about
such vulnerabilities through semantic reasoning.

\begin{findingbox}
\textbf{A4:}
\sys reproduced all three qualified FFmpeg 1-day CVEs with sanitizer-confirmed PoVs,
and discovered three previously unknown vulnerabilities in Bento4,
validating real-world applicability.
We confirmed these three vulnerabilities, and the maintainers have already fixed them.
\end{findingbox}

\section{Discussion and Future Work}
\label{sec:discuss}

\subsection{Comparison Among Baseline Fuzzers}
A notable observation from~\autoref{fig:fuzzer-comparison}
is that SOTA coverage-guided fuzzers (AFL++ w/ and w/o CmpLog)
significantly outperform directed greybox fuzzers (DGFs), and LLM-assisted fuzzers,
in both speed and vulnerability coverage.
While surprising, our results are consistent with recent studies
\cite{bao2025alarms,she2024fox,huang2023titan,geretto2025libaflgo},
which also report that modern coverage-guided fuzzers outperform DGFs.
These results seem to invalidate the underlying assumption that drives
a decade of directed fuzzing research:
\emph{distance-based guidance toward targets accelerates bug discovery}.

We believe such discrepancies stem from two main factors.
First, existing directed fuzzing approaches are fundamentally limited.
(i) Static analyses on low-level code structures (e.g., CFGs)
cannot prune the search space effectively to provide meaningful guidance.
(ii) Even if the guidance is perfect, random mutation cannot precisely follow the guidance.
Second, coverage-guided fuzzers essentially perform runtime path pruning:
uncovered paths are also paths that are more likely to reach the target.
Thus, general improvements in coverage-guided fuzzing,
like those in the AFL++ used in our experiments,
could have greater impact on PoV generation than distance-based guidance.

In contrast, \sys represents a paradigm shift in directed fuzzing:
Agentic reasoning can yield accurate and concise guidance,
and property-based testing can precisely follow the guidance to explore the search space.

\subsection{Future Directions for PoV Generation}
The design and evaluation of \sys were mainly focused on generating PoVs for
memory errors in popular C/C++ programs with existing fuzzing harnesses.
And for the Magma benchmark, the vulnerability triggering conditions are readily
available from the \cc{MAGMA\_LOG} call.
Therefore, several critical topics remain beyond the scope of this work
and warrant exploration in future research:
\begin{itemize}[leftmargin=*,topsep=0pt]
\item \textbf{Generalization to Other Programming Languages.}
  LLMs understand multiple programming languages, and \sys's constraint inference
  primarily relies on language-agnostic MCP tools (e.g., \cc{Glob}, \cc{Grep}, \cc{Read}).
  Thus, our approach is likely generalizable to other programming languages beyond C/C++.
\item \textbf{Generalization to Other Vulnerability Types.}
  Recent studies have shown that LLMs are also effective in
  generating static checkers from patches~\cite{yang2025knighter}.
  Thus, it is interesting to explore whether \sys can be generalized to
  other vulnerability types beyond memory safety bugs.
\item \textbf{Generalization to Other Input Formats.}
  In our evaluation, we only ask the agent to synthesize file input generators.
  Conceptually, LLMs should also be capable of generating other types of inputs,
  such as network/IPC communications (protocols) and user interactions.
\item \textbf{Agentic Harness Synthesis.}
  Recent studies have explored using LLMs to generate test harnesses~\cite{llmfuzzgen,lyu2024prompt},
  but not in an agentic manner.
  Although our workflow does not ask the agent to synthesize fuzzing harnesses,
  we observed it can switch to the right harness or generate new harness.
  Therefore, it is promising to explore agentic harness synthesis as part of future work.
\end{itemize}

\section{Related Work}
\label{sec:related}

The appendix~(\autoref{tab:related-comparison})
categorizes related work along four dimensions critical to PoV generation.

\PP{Traditional and LLM-Assisted Fuzzing}
Traditional directed greybox fuzzers guide exploration via
distance metrics~\cite{bohme2017directed,huang2022beacon,kim2023dafl}.
Recent LLM-assisted approaches leverage seed generation~\cite{deng2023large,xia2024fuzz4all}
and grammar extraction~\cite{meng2024large,zhang2025low}.
Yet all inherit the same fundamental limitation:
they rely on random mutation for constraint solving.

\PP{Agentic Directed Fuzzing and \sys Positioning}
Most LLM-assisted tools operate in one-shot mode,
generating artifacts once per target, then delegating to traditional fuzzers.
Locus~\cite{zhu2025locus} adds agentic predicate synthesis
but still relies on random mutation.
Sapia and B\"ohme~\cite{sapia2026reachabilitygap} employ agents for reachability
but target high-complexity functions via static metrics rather than
known vulnerabilities, failing for complex input structures
like CVE-2017-9047's nested XML DTD models.
FaultLine~\cite{nitin2025faultline} demonstrates PoV-focused agentic reasoning
but relies on direct LLM inference, struggling with
nested paths and compositional constraints.
ACE~\cite{luo2026agentic} couples agentic control with concolic execution
but optimizes coverage-oriented path exploration rather than
directed PoV generation.
Claude Security~\cite{anthropic2026claudesecurity} automates a
scan--validate--patch complete workflow:
parallel vulnerability discovery, multi-stage finding triage,
and remediation suggestions for human review.
\sys excludes repository-wide security auditing
and automatic patch generation.

\sys diverges fundamentally: we replace random mutation with
property-based parameterized generators encoding learned constraints.
Rather than tasking LLMs with slow direct input generation,
\sys synthesizes generators enabling high-throughput PBT exploration.
\sys leverages specialized MCP tools for fine-grained execution feedback,
enabling hypothesis validation beyond coarse coverage signals.
This agentic design addresses the semantic gap through
iterative feedback-driven reasoning and property-based constraint solving.

\section{Conclusion}
\label{sec:conclusion}
This paper presents \sys, an agentic framework for proof-of-vulnerability input generation.
\sys embodies a novel approach that leverages LLM agents to
(1) infer semantic level constraints required to trigger target vulnerabilities (PLAN),
(2) encode the extracted constraints as input-level parameterized spaces (IMPLEMENT),
(3) perform property-based testing to solve the constraints (EXECUTE), and
(4) use fine-grained execution feedback to iteratively refine hypotheses (REFLECT).
Our architecture combines on-demand tool orchestration,
persistent memory management, fine-grained execution feedback,
and property-based testing to bridge the semantic gap
between vulnerability comprehension and PoV generation.
Experimental evaluation on the Magma benchmark demonstrates that
\sys outperforms SOTA coverage-guided greybox fuzzers, directed greybox fuzzers and LLM-assisted fuzzers,
triggering 57 vulnerabilities with superior efficiency and consistency.
The results validate that LLMs can serve as autonomous reasoning engines
for complex security tasks beyond auxiliary code generation roles.

\section*{Acknowledgments}
We thank the anonymous reviewers for their insightful comments and valuable suggestions.
This work was supported, in part, by the National Science Foundation under Grant No. 2046026 and the Defense Advanced Research Projects Agency (DARPA) under Agreement No. HR00112590041.
Any opinions, findings, conclusions, or recommendations expressed in this paper are those of the authors and do not necessarily reflect the views of the funding agencies.

\section*{Ethical Considerations}
\label{sec:ethics}
\textbf{Stakeholders}: Software maintainers, end users, and the security community.
\textbf{Potential Harms}: Premature disclosure could enable exploitation before a fix exists;
the underlying technique could also be misused to automate vulnerability discovery for malicious ends.
\textbf{Mitigations}: We privately reported all findings to maintainers and withheld PoC details
and generator implementations until fixes shipped.
\textbf{Benefits}: Early, systematic vulnerability discovery enables timely patches
and advances defensive security research for the broader ecosystem.
\textbf{Decision Rationale}: Weighing security gains against exploitation risk,
we followed responsible disclosure to minimize harm while maximizing benefit,
consistent with the Menlo Report's ``Respect for Law and Public Interest'' principle.
Benchmark evaluation relies solely on previously disclosed CVEs, raising no additional concerns.

\section*{Generative AI Usage}

This paper was prepared with assistance from Cursor leveraging Sonnet 4.5 model.
Specifically, the authors utilized Cursor for:
(1) condensing the manuscript to meet the page limit while preserving technical accuracy and clarity;
(2) checking English grammar and ensuring compliance with the ACM sigconf format;
(3) implementing the MCP tools for the \sys prototype system.

All AI-assisted content was manually reviewed and validated by the authors.
The condensed text was verified to accurately reflect the original technical contributions.
The implemented MCP tools were validated through comprehensive unit tests to ensure correctness.

\bibliographystyle{ACM-Reference-Format}
\bibliography{conf,main}

\newpage
\section{Appendix}
\label{sec:appendix}
\subsection{Open Science}
\label{sec:open-science}

\subsubsection{Benchmarks}

We evaluate \sys on the Magma benchmark~\cite{hazimeh2020magma},
a ground-truth fuzzing suite with real-world CVE vulnerabilities.
Each target includes instrumentation via \cc{magma\_log} function calls
that distinguish vulnerability code execution from triggering.
The \cc{magma\_log} function emits diagnostic messages:
``MAGMA: Bug X reached'' indicates vulnerable code execution,
while ``MAGMA: Bug X triggered'' indicates condition satisfaction.
This separation enables precise measurement of PoV effectiveness.
We evaluate 129 CVEs and 9 security bugs across 9 projects:
\cc{libpng}, \cc{libtiff}, \cc{libxml2}, \cc{libsndfile}, \cc{poppler},
\cc{openssl}, \cc{php}, \cc{lua}, and \cc{sqlite3}.
All campaigns---\sys and every baseline---use Magma ISAN mode:
satisfied predicates raise \cc{SIGSEGV}.
Thus, every fuzzer receives an equivalent crash oracle
after satisfying the predicate, without runtime sanitizers.
\sys additionally consumes reach/no-trigger diagnostic signals
and predicate semantics during generation.
The observed gap therefore reflects algorithmic contribution,
not exclusive access to predicate-satisfaction crashes.

\subsubsection{Baseline Configuration}

We evaluate \sys against baseline approaches across three categories:

\begin{itemize}[leftmargin=1.5em]
  \item {\bf Directed Greybox Fuzzers}:
  AFLGo (commit \cc{fa125da}), first directed fuzzer with recent updates;
  SelectFuzz (docker \cc{11ff923}), SOTA open-sourced directed fuzzer
  considering control- and data-flow dependencies.
  
  \item {\bf Coverage-Guided Fuzzer}:
  AFL++ (release \cc{4.32c}) with cmpLog enabled and disabled.
  
  \item {\bf LLM-Assisted Fuzzer}:
  G2Fuzz (commit \cc{f62cc55}), SOTA chatbot-style LLM-assisted fuzzer.
  Since G2Fuzz lacks native Claude API support,
  we implemented this integration for fair comparison.
\end{itemize}

For RQ1, we incorporate published results from nine fuzzers (2022-2025):
MC2, Titan, Llamafuzz, AFLRun, SeedMind, LibAFLGo, Lyso, Locus, CSFuzz.
These fuzzers use 24-hour timeout across 10 trials (Llamafuzz: one month).

We exclude FaultLine~\cite{nitin2025faultline} and ACE~\cite{luo2026agentic}
from empirical baselines because their objectives diverge from ours.
FaultLine is CVE-report-driven, 
while ACE performs coverage-oriented agentic concolic exploration.
A direct comparison would require substantial adaptation to predicate-aware PoV generation 
and Magma's per-bug oracle evaluation, so we prefer not to overclaim without a fair setup. 

\subsubsection{Hardware and Environment Configuration}

Experiments execute on containerized environments:
server with 2x Intel Xeon E5-2695 v4 processors (36 cores) and 512GB RAM.
Each fuzzing instance: Docker container with 1 CPU core and 4GB RAM.
Baseline greybox fuzzers: non-deterministic mode enabled.
Baseline fuzzers: 24 hours per target using Magma's built-in seeds.
Baseline fuzzers: 10 repeated trials to capture randomness.

\sys configuration:
single run per target due to external LLM API budget constraints.
30-minute timeout per target for constraint refinement.
Cursor CLI version 2025.11.25-d5b3271 with Claude Sonnet-4.5.

G2Fuzz configuration:
24 hours per target, one trial run.
Claude Sonnet-4.5 API as the backbone LLM model.
Both G2Fuzz-generated seeds and Magma benchmark seeds.

\subsubsection{Static Analysis Cost and Target Selection}

Magma contains 1-5 fuzz harnesses per project (361 total targets).
Static analysis cost varies significantly:
libpng, libtiff, lua, libsndfile, sqlite3: within 10s per target.
poppler, php: 10-20 minutes per target.
openssl: 1-5 hours per target, several timing out.
To maintain feasibility, program analysis MCP tools disabled for OpenSSL
targets exceeding analysis time budget.
Reachability static analysis identifies 83 unreachable targets
in whole-program call graph.
Evaluation conducted on 278 reachable targets.

\subsubsection{Artifact Availability and Reproducibility}

The complete \sys implementation is available at:
\url{https://anonymous.4open.science/r/directed_property_based_fuzzer-2E1E}

This repository contains:
\begin{itemize}[leftmargin=1em]
\item Complete \sys source code and MCP tool implementations
\item \sys experimental evaluation scripts and harnesses
\item Documentation for reproduction and evaluation on Magma benchmark
\end{itemize}

The artifact is made available for reproducibility and artifact review
during the double-blind review process.
After publication, the complete implementation will be open-sourced
to enable long-term reproducibility and community adoption.

\subsection{FFmpeg CVE Selection}
\label{sec:app-ffmpeg-selection}

We audited all 2026 CVEs mentioning FFmpeg
to construct the 1-day reproduction target set.
The protocol first separated upstream FFmpeg vulnerabilities
from downstream applications, wrappers, and misattributed records.
It then retained only user-space bugs reachable through files
or media streams, controlled by structured input fields,
and described with component-level semantic detail.
Finally, we excluded candidates with public ready-to-run PoCs,
test cases, reproduction scripts, crash inputs,
or step-by-step reproduction artifacts.
Patches and advisory text were permitted,
because they provide semantic clues rather than executable PoCs.

\newcommand{\cveY}{$\checkmark$}
\newcommand{\cveN}{$\times$}
\newcommand{\cveP}{$\triangle$}
\newcommand{\cvePorN}{$\triangle$/$\times$}
\newcommand{\cveNA}{{\scriptsize---}}
\newcommand{\cveMeet}{\textbf{MEET}}
\newcommand{\cveDrop}{\textbf{DROP}}
\begin{table}[tbp]
\centering
\small
\setlength{\tabcolsep}{2.5pt}
\caption{%
FFmpeg CVE selection criteria.
\textbf{Criteria}: \textbf{Up.} upstream FFmpeg;
\textbf{Inp.} file or media stream;
\textbf{Str.} structured fields;
\textbf{Det.} semantic detail;
\textbf{PoC} no public ready-to-run PoC;
\textbf{Pat.} security patch available.
\textbf{Symbols}: \cveY\ satisfied;
\cveN\ not satisfied;
\cveP\ partial or unclear;
\cveNA\ not applicable.}
\label{tab:ffmpeg-cve-selection}
\begin{tabular}{@{}lcccccccc@{}}
\toprule
\textbf{CVE ID} &
\multicolumn{6}{c}{\textbf{Selection criteria}} &
\textbf{Outcome} \\
\cmidrule(lr){2-7}
& \textbf{Up.} & \textbf{Inp.} & \textbf{Str.} &
\textbf{Det.} & \textbf{PoC} & \textbf{Pat.} & \\
\midrule
\texttt{CVE-2026-40962} &
\cveY & \cveY & \cveY & \cveY & \cveY & \cveY &
\cveMeet \\
\texttt{CVE-2026-6385} &
\cveY & \cveY & \cveY & \cveY & \cveY & \cveN &
\cveMeet \\
\texttt{CVE-2026-30997} &
\cveY & \cveY & \cveY & \cveY & \cveY & \cveY &
\cveMeet \\
\texttt{CVE-2026-30998} &
\cveN & \cveN & \cveN & \cveP & \cveY & \cveP &
\cveDrop \\
\texttt{CVE-2026-30999} &
\cveP & \cveY & \cveN & \cveP & \cveY & \cveY &
\cveDrop \\
\texttt{CVE-2026-33024} &
\cveN & \cveN & \cveN & \cveNA & \cveNA & \cveY &
\cveDrop \\
\texttt{CVE-2026-35033} &
\cveN & \cveN & \cveN & \cveNA & \cveN & \cveY &
\cveDrop \\
\texttt{CVE-2026-35450} &
\cveN & \cveN & \cveN & \cveNA & \cveN & \cveP &
\cveDrop \\
\texttt{CVE-2026-3682} &
\cveN & \cveN & \cveN & \cveNA & \cveN & \cveP &
\cveDrop \\
\texttt{CVE-2026-22778} &
\cveN & \cveN & \cveN & \cveNA & \cveN & \cveY &
\cveDrop \\
\bottomrule
\end{tabular}
\end{table}

\subsection{Prompts}
\label{sec:app-prompts}

\begin{listing}[ht]
\caption{
Initial system prompt defining task objectives,
project configuration, and workflow rules.}
\label{lst:init-prompt}
\begin{lstlisting}[style=codefootnotesize]
# Security Research Autonomous Agent
## Mission
You are a security expert specializing in C/C++ vulnerabilities.
Your goal is to find a PoV input that can reproduce the target vulnerability.

## Configuration-Driven Workflow
**CRITICAL**: Your entire task is defined in two configuration files:
1. **`./project_config.md`**: Contains project info, goal, patterns, and technical details
2. **`./workflow_state.md`**: Contains workflow rules, phase transitions, and dynamic state

## Startup Protocol
1. **Read Configuration**: Always start by reading both configuration files
2. **Follow Workflow Rules**: All rules are defined in workflow_state.md STATIC sections

## Core Principle
Let the configuration files guide you completely.
Don't assume requirements - read and follow them exactly.
\end{lstlisting}
\end{listing}

\begin{listing}[ht]
\caption{
Initial system prompt template for cursor agent}
\label{lst:cursor-prompt}
\begin{lstlisting}[style=codefootnotesize]
## Mission
You are a security expert that finds PoV input for bugs in C/C++ programs.
Your goal is to find a PoV input that triggers the bug predicate.

## System Input
You will receive structured input containing:
1. **ENTRY_SOURCE_CODE**: Entry functions (main/FuzzerTestOneInput) and target location functions with line numbers
2. **TARGET_LOCATIONS**: Bug predicate locations with format 'loc=file:line, code=source_line'

Use this information to analyze the program, locate the bug predicate, and write a Python script that generates a PoV input which triggers the bug at the target location.
\end{lstlisting}
\end{listing}

\subsection{Markdown Templates}
\label{sec:app-markdown-templates}

\begin{listing}[ht]
\caption{
project\_config.md template.}
\label{lst:project-config}
\begin{lstlisting}[style=codefootnotesize]
# Project Configuration

<!-- STATIC:GOAL:START -->
## Goal
Find proof-of-concept inputs that violate safety properties in C/C++ programs using property-based directed fuzzing
<!-- STATIC:GOAL:END -->

<!-- STATIC:TOOLS_AND_REQUIREMENTS:START -->
## Available Tools
**Analysis MCP Tools**
- `get_callers`, `get_callees`: Call graph analysis
- `get_reaching_routes`: Routes and input files that reach targets
- `get_corpus_status`: Corpus analysis progress
- `extract_parameters`: Parameter space from reaching testcases
- `detect_deviation`: Find execution deviations from expected paths
- `get_generator_api_doc`: Generator API reference
- `fuzz`: Execute fuzzing with plan and generator
- `launch_gdb`: Launch interactive GDB session for advanced deviation analysis, root cause analysis, and TriggerPlan verification

**Workflow MCP Tools**
- `write_workflow_block(target_block, content_json)`: Write JSON to specific workflow blocks
- `transition_phase(next_phase)`: Transition to next phase with gatekeeper validation
- `check_phase_completion()`: Check if current phase tasks are completed
- `get_current_phase()`: Get current phase information
<!-- STATIC:TOOLS_AND_REQUIREMENTS:END -->

<!-- STATIC:TARGET_INFO:START -->
## Target Information
- **Binary**: {cmd}
- **Source Code**: {source_code_folder}
- **Output Directory**: {output_dir}
- **Reached Pattern**: {reached_pattern}
- **Triggered Pattern**: {triggered_pattern}
<!-- STATIC:TARGET_INFO:END -->

<!-- STATIC:SOURCE_CODE_BLOCKS:START -->
## Source Code Blocks
{source_code_blocks}
<!-- STATIC:SOURCE_CODE_BLOCKS:END -->

<!-- STATIC:TARGET_LOCATIONS:START -->
## Target Locations
{target_locations}
<!-- STATIC:TARGET_LOCATIONS:END -->

\end{lstlisting}
\end{listing}

\begin{lstlisting}[style=codefootnotesize]
# Workflow State
<!-- STATIC SECTIONS -->
<!-- STATIC:RULES:START -->
## Rules
- **MANDATORY**: Must enforce RULE_MANDATORY
- **Phase Gating**: Only allowed transitions:
```mermaid
graph LR
    PLAN --> IMPLEMENT
    IMPLEMENT --> EXECUTE
    EXECUTE --> REFLECT
    EXECUTE --> SUCCESS[PoC Found]
    REFLECT --> PLAN
```

### PLAN Phase Rules
- **R-PL1**: On first entry, must read project_config.md Source Code blocks (entry + target functions) and Target Locations
- **R-PL2**: Must write/update BugPredicates based on Target Locations
- **R-PL3**: Must write/update Preconditions, RootCauses, and TriggerPlans based on Source Code analysis and reflection insights
- **R-PL4**: Forbidden to perform any manual test. Must apply RULE_FLOW to verify your hypothesis.
- **R-PL5**: Must apply RULE_IMPLICIT_BEHAVIOR and RULE_MULTI_TARGETS
- **R-PL6**: ALLOWED TOOLS: get_callers, get_callees, get_reaching_routes, get_corpus_status, and Workflow MCP Tools

### IMPLEMENT Phase Rules
- **R-IM1**: Must take ALL TriggerPlans and convert input constraints into concrete ParameterSpace
- **R-IM2**: Must enumerate every possible way the bug condition can be met in ParameterSpace. Must grep and consider all other values for categorical parameters
- **R-IM3**: Must generate FuzzPlan with 5-10 concrete tests covering ALL TriggerPlans. Must apply RULE_POC_WINDOW
- **R-IM4**: Must define Breakpoints for validating preconditions and capturing runtime state at bug sites
- **R-IM5**: ALLOWED TOOLS: extract_parameters, get_generator_api_doc, and Workflow MCP Tools

### EXECUTE Phase Rules
- **R-EX1**: Must execute fuzz MCP tool using FuzzPlan and Breakpoints from IMPLEMENT phase
- **R-EX2**: Must update Metrics after fuzzing completes
- **R-EX3**: If bug triggered (pattern matched), must transition to SUCCESS
- **R-EX4**: If bug not triggered, must transition to REFLECT
- **R-EX5**: ALLOWED TOOLS: fuzz, get_generator_api_doc, and Workflow MCP Tools; other actions are forbidden

### REFLECT Phase Rules
- **R-RF1**: Must analyze why testcases in FuzzPlan failed to trigger the bug. Focus only on testcase failure analysis, not memory updates. Transition to PLAN when ready to update memory based on findings.
- **R-RF2**: For no-reach testcases in FuzzPlan, must use detect_deviation to identify which preconditions were not satisfied
- **R-RF3**: For reach/no-trigger testcases in FuzzPlan, must identify why bug predicate was not triggered by tracing variable dependencies backward
- **R-RF4**: Must transition to PLAN phase if performed more than THREE manual test. This budget resets upon re-entering REFLECT.
- **R-RF5**: ALLOWED TOOLS: detect_deviation, launch_gdb, get_callers, get_callees, and Workflow MCP Tools

### RULE_IMPLICIT_BEHAVIOR:
- Never assume explicit code paths are the only ones
- Always account for implicit library behavior, special cases and compatibility hacks.
- Perform broader related code reading across the codebase.

### GATEKEEPER Rules (STRICTLY ENFORCED)
- **G-1**: RULE_PHASE_GATING
- **G-2**: Memory data modification permissions: PLAN (BugPredicates, Preconditions, RootCauses, TriggerPlans), IMPLEMENT (ParameterSpace, FuzzPlan, Breakpoints), EXECUTE (Metrics, ParameterSpace), REFLECT (none - read-only)
- **G-3**: RULE_FLOW
- **G-4**: Auto-transition when phase tasks completed

### RULE_FLOW: PLAN→IMPLEMENT→EXECUTE→{REFLECT→PLAN | SUCCESS}
### RULE_POC_WINDOW: Prioritize malformed or boundary-skewed inputs that can bypass format checks to trigger bugs, not fully valid ones.
### RULE_MANDATORY: Always read workflow_state.md before every phase transition. Use workflow MCP server tools to update it.
### RULE_FILE_OPS: Must use workflow MCP server tools (write_workflow_block, transition_phase) for safe reading/writing
### RULE_SAFE_UPDATE: Never overwrite whole blocks unintentionally.
### RULE_PHASE_GATING: Each MCP tool blocked unless in correct phase with required prerequisites
### RULE_MEMORY: All state persisted in workflow_state.md; no ephemeral memory allowed
### RULE_DOCS: Do not create any extra markdown document other than workflow_state.md and project_config.md
### RULE_MAGMA: Do not analyze Magma benchmark instrumentation. See magma.md.
### RULE_FUZZ_TOOL: Only fuzz MCP tool in EXECUTE phase can declare PoC
### RULE_MULTI_TARGETS: Triggering one target is sufficient. If a target has multiple triggering conditions, satisfy any bug predicate is sufficient. Prioritize simpler bug predicates.
<!-- STATIC:RULES:END -->

<!-- DYNAMIC SECTIONS -->
<!-- These sections are managed by the AI during workflow execution -->
<!-- DYNAMIC:STATE:START -->
## State
```json
{
  "phase": "PLAN",
  "status": "Starting directed fuzzing workflow",
  "current_task": "Analyze target and create initial plan",
  "next_action": "Read project_config.md and extract BugPredicates"
}
```
<!-- DYNAMIC:STATE:END -->

<!-- DYNAMIC:BUG_PREDICATES:START -->
## BugPredicates
```json
[]
```
<!-- DYNAMIC:BUG_PREDICATES:END -->

<!-- DYNAMIC:PRECONDITIONS:START -->
## Preconditions
```json
[]
```
<!-- DYNAMIC:PRECONDITIONS:END -->

<!-- DYNAMIC:ROOT_CAUSES:START -->
## RootCauses
```json
[]
```
<!-- DYNAMIC:ROOT_CAUSES:END -->

<!-- DYNAMIC:PARAMETER_SPACE:START -->
## ParameterSpace
```json
{}
```
<!-- DYNAMIC:PARAMETER_SPACE:END -->

<!-- DYNAMIC:TRIGGER_PLANS:START -->
## TriggerPlans
```json
[]
```
<!-- DYNAMIC:TRIGGER_PLANS:END -->

<!-- DYNAMIC:FUZZ_PLAN:START -->
## FuzzPlan
```json
[]
```
<!-- DYNAMIC:FUZZ_PLAN:END -->

<!-- DYNAMIC:BREAKPOINTS:START -->
## Breakpoints
```json
[]
```
<!-- DYNAMIC:BREAKPOINTS:END -->

<!-- DYNAMIC:METRICS:START -->
## Metrics
```json
{
  "total_iterations": 0,
  "total_reached_count": 0,
  "last_reached_count": 0,
  "triggered_count": 0,
  "timeout_count": 0,
  "error_count": 0,
  "last_updated": ""
}
```
<!-- DYNAMIC:METRICS:END -->
\end{lstlisting}
\noindent\textbf{Listing 4: workflow\_state.md template.}
\label{lst:workflow-state}

\begin{table*}[!htbp]
\centering
\small
\caption{Comparison of fuzzing approaches for PoV input generation.}
\label{tab:related-comparison}
\begin{tabular}{@{}llcccc@{}}
\toprule
\textbf{Category} & \textbf{Approach} & \textbf{Goal} & \textbf{LLM Mode} & \textbf{Input Gen} & \textbf{Feedback} \\
\midrule
\multirow{2}{*}{Traditional DGF} 
& AFLGo~\cite{bohme2017directed}, Hawkeye~\cite{chen2018hawkeye} & Directed & -- & Mutation & distance \\
& Beacon~\cite{huang2022beacon}, DAFL~\cite{kim2023dafl} & Directed & -- & Mutation & distance \\
\midrule
\multirow{3}{*}{LLM-Coverage} 
& TitanFuzz~\cite{deng2023large}, Fuzz4All~\cite{xia2024fuzz4all} & Coverage & One-shot & LLM-Direct & Coverage \\
& SeedMind~\cite{shi2024harnessing}, ChatAFL~\cite{meng2024large} & Coverage & Iterative & Generator & Coverage \\
& G2Fuzz~\cite{zhang2025low}, HLPFUZZ~\cite{yang2025hybrid} & Coverage & Iterative & Generator & Coverage \\
\midrule
\multirow{2}{*}{LLM-Directed} 
& HGFuzzer~\cite{xu2025directed}, RANDLUZZ~\cite{feng2025fuzzing} & Directed & One-shot & Mutation & distance \\
& Locus~\cite{zhu2025locus} & Directed & Agentic & Mutation & distance \\
\midrule
\multirow{2}{*}{LLM-Symbolic} 
& AutoBug~\cite{li2025large}, COTTONTAIL~\cite{tu2025large} & Coverage & Iterative & LLM-Direct & Coverage \\
& CONCOLLMIC~\cite{luo2026agentic} & Coverage & Agentic & LLM-Direct & Coverage \\
\midrule
\multirow{2}{*}{Agentic} 
& Sapia \& B\"ohme~\cite{sapia2026reachabilitygap} & Coverage & Agentic & Mutation & Coverage \\
& FaultLine~\cite{nitin2025faultline} & PoV & Agentic & LLM-Direct & exit code \\
\midrule
& \textbf{\sys} & \textbf{PoV} & \textbf{Agentic} & \textbf{PBT Generator} & \textbf{Fine-grained} \\
\bottomrule
\end{tabular}
\end{table*}

\begin{table*}[!htbp]
\centering
\caption{MCP Tools invoked by \sys agent.}
\label{tab:mcp-tools}
\small
\begin{tabular}{@{}llp{4.5cm}p{5.5cm}@{}}
\toprule
\textbf{Phase} & \textbf{Tool} & \textbf{Signature} & \textbf{Description} \\
\midrule
ALL & \cc{write\_workflow\_block} & \cc{(target, content)} & Persists state updates \\
ALL & \cc{transition\_phase} & \cc{(next\_phase: str)} & Validates and enforces phase transitions \\
ALL & \cc{check\_phase\_completion} & \cc{()} & Queries prerequisites for current phase transition \\
\midrule
PLAN & \cc{get\_callers} & \cc{(function\_name: str)} & Retrieves call graph callers for specified function \\
PLAN & \cc{get\_callees} & \cc{(function\_name: str)} & Retrieves call graph callees for specified function \\
PLAN & \cc{get\_reaching\_routes} & \cc{()} & Returns callstacks and testcases reaching target locations \\
PLAN & \cc{get\_corpus\_status} & \cc{()} & Reports corpus processing progress \\
\midrule
IMPLEMENT & \cc{extract\_parameters} & \cc{(extractor\_code: str)} & Derives typed parameter specifications from reaching corpus \\
IMPLEMENT & \cc{get\_generator\_api\_doc} & \cc{(topic?: str)} & Returns generator API documentation and examples \\
\midrule
EXECUTE & \cc{fuzz} & \cc{(setting: dict)} & Executes property-based fuzzing with two-phase search \\
\midrule
REFLECT & \cc{detect\_deviation} & \cc{(input\_file\_path: str)} & Identifies violated reachability constraints via trace analysis \\
REFLECT & \cc{launch\_gdb} & \cc{(cmd: list[str], timeout: str)} & Spawns interactive debugger for runtime state inspection \\
\bottomrule
\end{tabular}
\end{table*}

\subsection{Supplementary Evaluation Results}
\label{sec:app-supplementary-eval}

\subsubsection{Time-to-Reach (TTR) Analysis}
\label{sec:app-ttr}

\autoref{tab:ttr} presents the Time-to-Reach (TTR) between different fuzzers.
As mentioned earlier, some vulnerabilities, or \cc{magma\_log}, are reachable by the provided initial seeds,
therefore we excluded those vulnerabilities from the comparison.
The results demonstrate that \sys reaches vulnerabilities significantly faster than other fuzzers on most benchmarks
(only 2 are slower and 2 never reached).
Compared to TTE (\autoref{tab:trigger}), it is evident that most vulnerabilities reached by \sys are triggered shortly thereafter,
except for \cc{SQL013}, \cc{SQL014}, \cc{TIF001}, and \cc{TIF002}.
Other fuzzers generally require significantly more time to transit from reaching to triggering the vulnerabilities,
or never managed to trigger within the time limit.

\begin{table}[!htbp]
\caption{
Time-to-Reach (TTR) comparison across benchmarks.
TTR denotes the time required to reach each vulnerability.
T.O denotes timeout at 24 hours. NA denotes not applicable, when the fuzzer cannot handle the target programs.
We excluded vulnerabilities reachable by initial benchmark seeds or cannot be reached by all tested fuzzers.
}
\label{tab:ttr}
{\footnotesize
\begin{tabular}{l|r rr rr rr rr}
\toprule
\textbf{Bug ID} & \textbf{PBFuzz} & \textbf{G2Fuzz} & \multicolumn{1}{c}{\textbf{AFL++}} & \multicolumn{1}{c}{\textbf{cmplog}} & \multicolumn{1}{c}{\textbf{AFLGo}} & \multicolumn{1}{c}{\textbf{SelectFuzz}} \\
\midrule
LUA001 & \cellcolor[HTML]{165128} 2.1m & \cellcolor[HTML]{FF9999} T.O & \cellcolor[HTML]{FF9999} T.O & \cellcolor[HTML]{60b472} 19.2h & \cellcolor[HTML]{FF9999} T.O & \cellcolor[HTML]{FF9999} T.O \\
LUA002 & \cellcolor[HTML]{60b472} 15s & \cellcolor[HTML]{165128} 6s & \cellcolor[HTML]{8ac794} 5.3h & \cellcolor[HTML]{b2dab7} 12.3h & \cellcolor[HTML]{d8edda} 15.1h & \cellcolor[HTML]{ffffff} 19.6h \\
LUA003 & \cellcolor[HTML]{60b472} 15s & \cellcolor[HTML]{165128} 6s & \cellcolor[HTML]{FF9999} T.O & \cellcolor[HTML]{FF9999} T.O & \cellcolor[HTML]{FF9999} T.O & \cellcolor[HTML]{FF9999} T.O \\
LUA004 & \cellcolor[HTML]{60b472} 17s & \cellcolor[HTML]{165128} 5s & \cellcolor[HTML]{8ac794} 3.6h & \cellcolor[HTML]{d8edda} 7.2h & \cellcolor[HTML]{ffffff} 8.5h & \cellcolor[HTML]{b2dab7} 4.7h \\
PDF002 & \cellcolor[HTML]{60b472} 18s & \cellcolor[HTML]{b2dab7} 5.2h & \cellcolor[HTML]{ffffff} 15.7h & \cellcolor[HTML]{d8edda} 14.4h & \cellcolor[HTML]{8ac794} 3.8h & \cellcolor[HTML]{165128} 17s \\
PDF004 & \cellcolor[HTML]{FF9999} T.O & \cellcolor[HTML]{165128} 37.3m & \cellcolor[HTML]{60b472} 12.5h & \cellcolor[HTML]{8ac794} 17.1h & \cellcolor[HTML]{b2dab7} 22.8h & \cellcolor[HTML]{FF9999} T.O \\
PDF005 & \cellcolor[HTML]{165128} 11s & \cellcolor[HTML]{8ac794} 5.2h & \cellcolor[HTML]{d8edda} 16.8h & \cellcolor[HTML]{b2dab7} 14.4h & \cellcolor[HTML]{ffffff} 21.6h & \cellcolor[HTML]{60b472} 17s \\
PDF008 & \cellcolor[HTML]{165128} 54s & \cellcolor[HTML]{FF9999} T.O & \cellcolor[HTML]{FF9999} T.O & \cellcolor[HTML]{FF9999} T.O & \cellcolor[HTML]{FF9999} T.O & \cellcolor[HTML]{FF9999} T.O \\
PDF018 & \cellcolor[HTML]{165128} 3.1m & \cellcolor[HTML]{b2dab7} 3.6h & \cellcolor[HTML]{d8edda} 10.7h & \cellcolor[HTML]{ffffff} 22.5h & \cellcolor[HTML]{8ac794} 21.8m & \cellcolor[HTML]{60b472} 10.6m \\
PHP001 & \cellcolor[HTML]{165128} 11m & \cellcolor[HTML]{FF9999} T.O & \cellcolor[HTML]{FF9999} T.O & \cellcolor[HTML]{FF9999} T.O & \cellcolor[HTML]{FF9999} T.O & \cellcolor[HTML]{FF9999} NA \\
PHP010 & \cellcolor[HTML]{165128} 19.6m & \cellcolor[HTML]{FF9999} T.O & \cellcolor[HTML]{FF9999} T.O & \cellcolor[HTML]{FF9999} T.O & \cellcolor[HTML]{FF9999} T.O & \cellcolor[HTML]{FF9999} NA \\
PHP013 & \cellcolor[HTML]{165128} 6.7m & \cellcolor[HTML]{FF9999} T.O & \cellcolor[HTML]{FF9999} T.O & \cellcolor[HTML]{FF9999} T.O & \cellcolor[HTML]{FF9999} T.O & \cellcolor[HTML]{FF9999} NA \\
PHP014 & \cellcolor[HTML]{165128} 3.9m & \cellcolor[HTML]{FF9999} T.O & \cellcolor[HTML]{FF9999} T.O & \cellcolor[HTML]{FF9999} T.O & \cellcolor[HTML]{FF9999} T.O & \cellcolor[HTML]{FF9999} NA \\
SND017 & \cellcolor[HTML]{165128} 7s & \cellcolor[HTML]{60b472} 29s & \cellcolor[HTML]{ffffff} 72.8m & \cellcolor[HTML]{b2dab7} 53.3m & \cellcolor[HTML]{8ac794} 17.9m & \cellcolor[HTML]{d8edda} 59.1m \\
SND020 & \cellcolor[HTML]{165128} 2m & \cellcolor[HTML]{ffffff} 4.8h & \cellcolor[HTML]{b2dab7} 82m & \cellcolor[HTML]{60b472} 66.7m & \cellcolor[HTML]{8ac794} 80.4m & \cellcolor[HTML]{d8edda} 99.1m \\
SQL001 & \cellcolor[HTML]{165128} 2.5m & \cellcolor[HTML]{FF9999} T.O & \cellcolor[HTML]{FF9999} T.O & \cellcolor[HTML]{FF9999} T.O & \cellcolor[HTML]{FF9999} T.O & \cellcolor[HTML]{FF9999} T.O \\
SQL003 & \cellcolor[HTML]{165128} 2.4m & \cellcolor[HTML]{FF9999} T.O & \cellcolor[HTML]{60b472} 15.2h & \cellcolor[HTML]{b2dab7} 20.7h & \cellcolor[HTML]{8ac794} 19.8h & \cellcolor[HTML]{FF9999} T.O \\
SQL006 & \cellcolor[HTML]{165128} 2.2m & \cellcolor[HTML]{FF9999} T.O & \cellcolor[HTML]{8ac794} 11.2h & \cellcolor[HTML]{60b472} 9.9h & \cellcolor[HTML]{b2dab7} 13h & \cellcolor[HTML]{d8edda} 14h \\
SQL009 & \cellcolor[HTML]{165128} 3.7m & \cellcolor[HTML]{ffffff} 7.4h & \cellcolor[HTML]{60b472} 3.9m & \cellcolor[HTML]{8ac794} 7.1m & \cellcolor[HTML]{d8edda} 23.3m & \cellcolor[HTML]{b2dab7} 12.1m \\
SQL011 & \cellcolor[HTML]{FF9999} T.O & \cellcolor[HTML]{FF9999} T.O & \cellcolor[HTML]{165128} 20.9h & \cellcolor[HTML]{FF9999} T.O & \cellcolor[HTML]{FF9999} T.O & \cellcolor[HTML]{60b472} 22.4h \\
SQL012 & \cellcolor[HTML]{165128} 1.8m & \cellcolor[HTML]{d8edda} 14h & \cellcolor[HTML]{60b472} 6h & \cellcolor[HTML]{8ac794} 10.3h & \cellcolor[HTML]{b2dab7} 10.9h & \cellcolor[HTML]{ffffff} 15.6h \\
SQL013 & \cellcolor[HTML]{165128} 11.4m & \cellcolor[HTML]{FF9999} T.O & \cellcolor[HTML]{8ac794} 20.6h & \cellcolor[HTML]{60b472} 17.7h & \cellcolor[HTML]{b2dab7} 20.9h & \cellcolor[HTML]{FF9999} T.O \\
SQL014 & \cellcolor[HTML]{165128} 3.1m & \cellcolor[HTML]{ffffff} 6.8h & \cellcolor[HTML]{60b472} 5.4m & \cellcolor[HTML]{8ac794} 16.6m & \cellcolor[HTML]{d8edda} 62.6m & \cellcolor[HTML]{b2dab7} 45m \\
SQL020 & \cellcolor[HTML]{165128} 2.1m & \cellcolor[HTML]{FF9999} T.O & \cellcolor[HTML]{60b472} 11.5h & \cellcolor[HTML]{8ac794} 15.8h & \cellcolor[HTML]{b2dab7} 17.7h & \cellcolor[HTML]{d8edda} 17.8h \\
SSL006 & \cellcolor[HTML]{165128} 22.4m & \cellcolor[HTML]{FF9999} T.O & \cellcolor[HTML]{FF9999} T.O & \cellcolor[HTML]{FF9999} T.O & \cellcolor[HTML]{FF9999} T.O & \cellcolor[HTML]{FF9999} NA \\
TIF001 & \cellcolor[HTML]{60b472} 10.8m & \cellcolor[HTML]{165128} 6s & \cellcolor[HTML]{8ac794} 7.9h & \cellcolor[HTML]{b2dab7} 17.5h & \cellcolor[HTML]{ffffff} 23.5h & \cellcolor[HTML]{d8edda} 18.4h \\
TIF002 & \cellcolor[HTML]{165128} 8.8m & \cellcolor[HTML]{8ac794} 2.5h & \cellcolor[HTML]{b2dab7} 4.2h & \cellcolor[HTML]{60b472} 2.4h & \cellcolor[HTML]{d8edda} 18.4h & \cellcolor[HTML]{ffffff} 19.8h \\
TIF005 & \cellcolor[HTML]{165128} 9.7m & \cellcolor[HTML]{8ac794} 82.3m & \cellcolor[HTML]{FF9999} T.O & \cellcolor[HTML]{60b472} 37.1m & \cellcolor[HTML]{FF9999} T.O & \cellcolor[HTML]{FF9999} T.O \\
TIF006 & \cellcolor[HTML]{165128} 1.9m & \cellcolor[HTML]{8ac794} 82.3m & \cellcolor[HTML]{b2dab7} 15.5h & \cellcolor[HTML]{60b472} 44.1m & \cellcolor[HTML]{ffffff} 22.3h & \cellcolor[HTML]{d8edda} 20.6h \\
TIF008 & \cellcolor[HTML]{165128} 3.2m & \cellcolor[HTML]{FF9999} T.O & \cellcolor[HTML]{8ac794} 14.1h & \cellcolor[HTML]{60b472} 13.8h & \cellcolor[HTML]{b2dab7} 19.9h & \cellcolor[HTML]{FF9999} T.O \\
TIF009 & \cellcolor[HTML]{165128} 4m & \cellcolor[HTML]{FF9999} T.O & \cellcolor[HTML]{8ac794} 4.7h & \cellcolor[HTML]{b2dab7} 6.2h & \cellcolor[HTML]{d8edda} 16.9h & \cellcolor[HTML]{60b472} 2.3h \\
TIF010 & \cellcolor[HTML]{165128} 4.9m & \cellcolor[HTML]{d8edda} 2.6h & \cellcolor[HTML]{8ac794} 12m & \cellcolor[HTML]{60b472} 6.1m & \cellcolor[HTML]{b2dab7} 17.9m & \cellcolor[HTML]{ffffff} 6.9h \\
XML002 & \cellcolor[HTML]{165128} 1.8m & \cellcolor[HTML]{FF9999} T.O & \cellcolor[HTML]{60b472} 17.7h & \cellcolor[HTML]{8ac794} 22.9h & \cellcolor[HTML]{FF9999} T.O & \cellcolor[HTML]{FF9999} T.O \\
XML010 & \cellcolor[HTML]{165128} 8.5m & \cellcolor[HTML]{FF9999} T.O & \cellcolor[HTML]{FF9999} T.O & \cellcolor[HTML]{FF9999} T.O & \cellcolor[HTML]{FF9999} T.O & \cellcolor[HTML]{FF9999} T.O \\
\bottomrule
\end{tabular}
}
\end{table}

\subsubsection{Reproducibility and Consistency Analysis}
\label{sec:app-consistency}

We evaluate reproducibility via consistency analysis across repeated executions.
Due to cost concern, we selected 23 vulnerabilities that all fuzzers could trigger at least once.
Each system executes 10 independent runs per vulnerability.

\begin{figure}[!htbp]
\centering
\includegraphics[width=0.9\columnwidth,height=0.5\textheight]{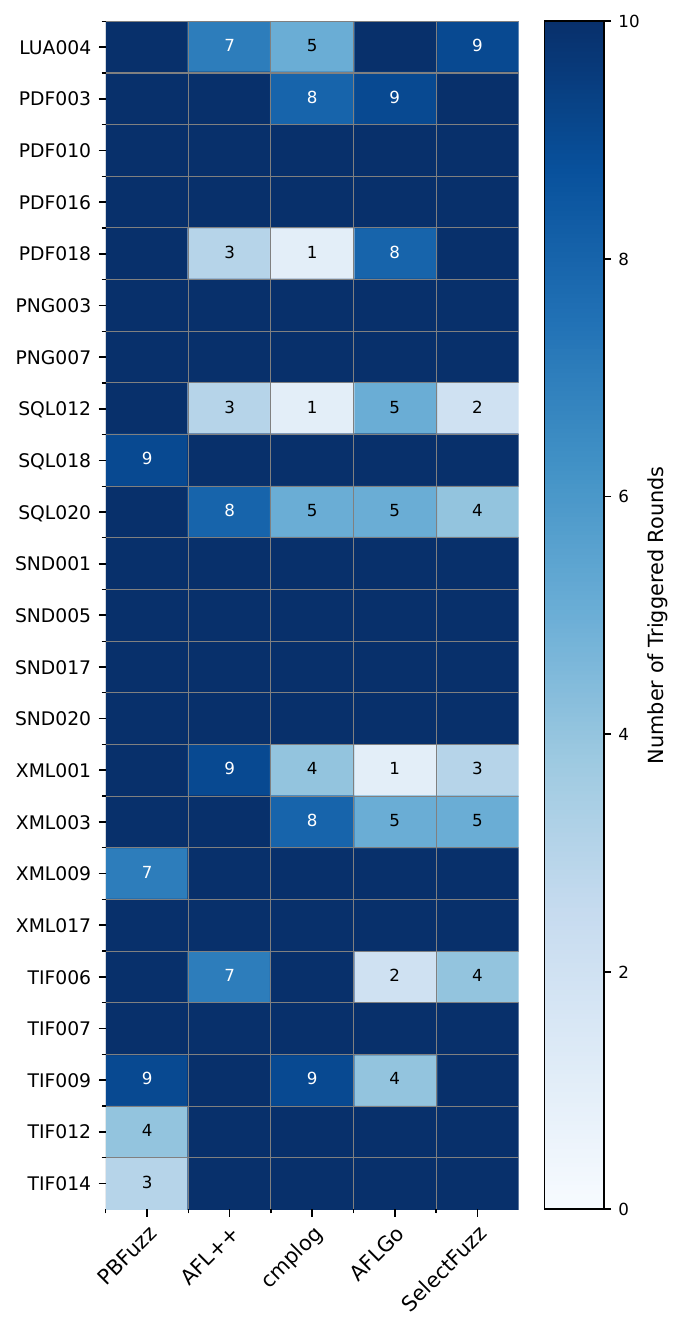}
\caption{Reproducibility heatmap: triggering frequency across runs.
Vertical axis lists vulnerabilities, horizontal axis lists fuzzers.
Numeric annotations highlight instances below maximum.}
\label{fig:consistency-heatmap}
\end{figure}

\autoref{fig:consistency-heatmap} summarizes consistency results.
\sys achieves superior consistency attributed to its deterministic constraint solving mechanism.
Empirically, \sys triggers 18 of 23 tested vulnerabilities consistently (10/10 runs),
including SQL012 and PDF018 where AFL++ and cmplog achieve only 3/10 and 1/10 respectively.
This result further highlights the effectiveness of \sys's novel approach,
which leverages agentic reasoning to directly synthesize parameterized generators.
This significantly improves the likelihood that generated inputs satisfy
reachability preconditions by construction.
As a result, \sys not only can trigger the vulnerabilities faster,
but also does so more consistently across repeated runs.

\subsubsection{Model Variant Analysis}
\label{sec:app-model-variants}

To isolate the contribution of architectural design
from inherent model capabilities,
we evaluated four distinct LLM variants
on representative benchmarks spanning diverse input formats.
We selected two challenging programs---SQLite (text-based SQL queries)
and LibTIFF (binary format)---to evaluate generalization across format diversity.
Each model variant executed a single trial per vulnerability with 30-minute timeout.

\begin{table}[!htbp]
\centering
\small
\caption{Model variant TTE comparison: Sonnet 4.5 vs.
extended reasoning vs. GPT-5 vs. Grok on SQLite and LibTIFF.}
\label{tab:model-variants}
\begin{tabular}{l|rrrr}
\toprule
\textbf{Bug ID} & \textbf{Sonnet-4.5} &
\textbf{4.5-Thinking} & \textbf{GPT-5} & \textbf{Grok-4} \\
\midrule
SQL001 & T.O & T.O & T.O & T.O \\
SQL002 & T.O & T.O & 627 & T.O \\
SQL003 & 143 & 279 & 216 & T.O \\
SQL006 & T.O & T.O & T.O & T.O \\
SQL007 & 339 & T.O & T.O & T.O \\
SQL009 & T.O & T.O & T.O & T.O \\
SQL010 & 334 & T.O & 906 & T.O \\
SQL011 & T.O & 1058 & T.O & T.O \\
SQL012 & 120 & 320 & 776 & T.O \\
SQL013 & T.O & T.O & T.O & T.O \\
SQL014 & T.O & T.O & T.O & T.O \\
SQL015 & 493 & 320 & 226 & 208 \\
SQL016 & T.O & T.O & T.O & T.O \\
SQL017 & T.O & T.O & T.O & T.O \\
SQL018 & 208 & 378 & 217 & T.O \\
SQL019 & T.O & T.O & T.O & T.O \\
SQL020 & 127 & 303 & T.O & 334 \\
TIF001 & T.O & T.O & T.O & T.O \\
TIF002 & T.O & T.O & T.O & 158 \\
TIF003 & T.O & T.O & T.O & T.O \\
TIF005 & 579 & 433 & 414 & T.O \\
TIF006 & 113 & 360 & 973 & T.O \\
TIF007 & 411 & 384 & 326 & 669 \\
TIF008 & 457 & 310 & 303 & T.O \\
TIF009 & 1214 & 355 & 295 & T.O \\
TIF010 & T.O & T.O & T.O & T.O \\
TIF012 & 1178 & T.O & T.O & T.O \\
TIF014 & 767 & T.O & 292 & T.O \\
\bottomrule
\end{tabular}
\end{table}

\autoref{tab:model-variants} quantifies model variant performance.
Sonnet-4.5 triggers 14 vulnerabilities,
outperforming 4.5-Thinking (10 vulnerabilities), GPT-5 (11 vulnerabilities), and Grok-4 (4 vulnerabilities).
This ranking reveals some insights about agent architecture and model capabilities.

First, Sonnet-4.5 outperforms both GPT-5 and Grok-4.
We attribute this to Claude's focus on coding tasks.
We argue that, akin to human experts, superior code reasoning capabilities enable a model to
(1) understand program semantics to infer constraints,
(2) trim the search space for efficient exploration, and
(3) synthesize correct generators to solve constraints.
Concretely, on SQL020, Sonnet-4.5 identified the critical parameter
\cc{bIntToNull=1} in window function processing via call graph analysis.
This semantic insight enabled focused generator synthesis with correct
harness protocol enforcement, achieving 127s whereas GPT-5 timed out.
This result highlights the scalability of \sys's architecture---as LLMs improve,
\sys's ability to generate PoV will correspondingly improve.

The second interesting observation is that Sonnet-4.5
outperforms its extended reasoning variant (4.5-Thinking).
We attribute this to \sys's workflow design.
PoV generation is inherently a multi-phase complex task.
\sys decomposes this complex task based on human expertises into
structured workflow phases with persistent memory maintaining reasoning state.
When facing such tasks, an LLM's internal inference time reasoning may
overthink and hallucinate~\cite{aggarwal2025optimalthinkingbench, yao2025reasoningmodelspronehallucination};
while the non-thinking mode can stick to the workflow structure and rely on
human expertise to guide the reasoning process.
For instance, for SQL003, Sonnet-4.5 succeeded in 143s
while 4.5-Thinking requires 279s---excessive internal reasoning
consumes computational budgets without accelerating convergence.
This result demonstrates that for LLM-based agents,
architecture-level workflow design with task-specific (e.g., PoV generation)
decomposition can be more effective and efficient than model-level reasoning.

We also examined Grok-4's poor performance and found
a distinct bottleneck: inadequate MCP tool support.
While Sonnet-4.5 and GPT-5 can invoke tools efficiently,
Grok-4 struggles to satisfy MCP input schema constraints,
expending computational effort formatting tool parameters
rather than performing semantic constraint inference.
This architectural incompatibility cascades: tool call failures
require retries that consume remaining time budgets.
This result highlights two insights:
(1) effective agent-tool integration is critical for complex tasks, and
(2) for agentic architectures, model-tool fit matters.

\subsubsection{Stage-Level PIER Ablation Data}
\label{sec:app-stage-level-data}

\begin{table*}[!htbp]
\centering
\small
\renewcommand{\arraystretch}{0.96}
\setlength{\tabcolsep}{3.0pt}
\caption{%
Per-round stage-level PIER ablation data
for the $11$ targets in \autoref{fig:stage-level-ablation}
($49$ PIER rounds total).
\textbf{Raw counts:}
GT\# is the number of ground-truth constraints;
Pred\# is the number of agent-inferred constraints;
M\_P\# is the number of predicted constraints matching GT;
M\_G\# is the number of GT constraints covered;
Impl\# is the number of inferred constraints parameterized;
Faith\# is the number of parameterized constraints correctly encoded.
\textbf{Metrics:}
Prec $=$ M\_P\#/Pred\#;
Rec $=$ M\_G\#/GT\#;
Cov $=$ Impl\#/Pred\#;
Fid $=$ Faith\#/Impl\#.
Reach and Trig are binary outcomes
($p_{\mathit{reach}}$ and $p_{\mathit{trigger}}$).}
\label{tab:stage-level-data}
\begin{tabular}{@{}l r rrrr rr rr rr rr@{}}
\toprule
& & \multicolumn{4}{c}{\textbf{PLAN (raw)}}
& \multicolumn{2}{c}{\textbf{IMPLEMENT (raw)}}
& \multicolumn{2}{c}{\textbf{PLAN (metric)}}
& \multicolumn{2}{c}{\textbf{IMPLEMENT (metric)}}
& \multicolumn{2}{c@{}}{\textbf{EXECUTE}} \\
\cmidrule(lr){3-6} \cmidrule(lr){7-8}
\cmidrule(lr){9-10} \cmidrule(lr){11-12} \cmidrule(l){13-14}
\textbf{Bug} & \textbf{Rnd}
& \textbf{GT\#} & \textbf{Pred\#} & \textbf{M\_P\#} & \textbf{M\_G\#}
& \textbf{Impl\#} & \textbf{Faith\#}
& \textbf{Prec} & \textbf{Rec}
& \textbf{Cov} & \textbf{Fid}
& \textbf{Reach} & \textbf{Trig} \\
\midrule
\multirow{8}{*}{\cc{PDF005}} & 1 & 5 & 4 & 4 & 4 & 2 & 2 & 1.00 & 0.80 & 0.50 & 1.00 & 0 & 0 \\
 & 2 & 5 & 4 & 4 & 4 & 2 & 2 & 1.00 & 0.80 & 0.50 & 1.00 & 0 & 0 \\
 & 3 & 5 & 4 & 4 & 4 & 2 & 2 & 1.00 & 0.80 & 0.50 & 1.00 & 0 & 0 \\
 & 4 & 5 & 4 & 4 & 4 & 2 & 2 & 1.00 & 0.80 & 0.50 & 1.00 & 0 & 0 \\
 & 5 & 5 & 4 & 4 & 4 & 2 & 2 & 1.00 & 0.80 & 0.50 & 1.00 & 0 & 0 \\
 & 6 & 5 & 4 & 4 & 4 & 2 & 2 & 1.00 & 0.80 & 0.50 & 1.00 & 0 & 0 \\
 & 7 & 5 & 4 & 4 & 4 & 2 & 2 & 1.00 & 0.80 & 0.50 & 1.00 & 0 & 0 \\
 & 8 & 5 & 4 & 4 & 4 & 2 & 2 & 1.00 & 0.80 & 0.50 & 1.00 & 1 & 1 \\
\midrule
\multirow{5}{*}{\cc{PDF012}} & 1 & 4 & 3 & 3 & 3 & 2 & 1 & 1.00 & 0.75 & 0.67 & 0.50 & 1 & 0 \\
 & 2 & 4 & 4 & 4 & 4 & 3 & 1 & 1.00 & 1.00 & 0.75 & 0.33 & 1 & 0 \\
 & 3 & 4 & 4 & 4 & 4 & 4 & 2 & 1.00 & 1.00 & 1.00 & 0.50 & 1 & 0 \\
 & 4 & 4 & 4 & 4 & 4 & 4 & 2 & 1.00 & 1.00 & 1.00 & 0.50 & 1 & 0 \\
 & 5 & 4 & 5 & 4 & 4 & 4 & 4 & 0.80 & 1.00 & 0.80 & 1.00 & 1 & 1 \\
\midrule
\multirow{7}{*}{\cc{PDF019}} & 1 & 4 & 2 & 2 & 3 & 2 & 2 & 1.00 & 0.75 & 1.00 & 1.00 & 0 & 0 \\
 & 2 & 4 & 2 & 2 & 3 & 2 & 2 & 1.00 & 0.75 & 1.00 & 1.00 & 0 & 0 \\
 & 3 & 4 & 1 & 1 & 1 & 1 & 1 & 1.00 & 0.25 & 1.00 & 1.00 & 1 & 0 \\
 & 4 & 4 & 2 & 1 & 1 & 2 & 1 & 0.50 & 0.25 & 1.00 & 0.50 & 1 & 0 \\
 & 5 & 4 & 2 & 2 & 3 & 2 & 2 & 1.00 & 0.75 & 1.00 & 1.00 & 1 & 0 \\
 & 6 & 4 & 2 & 2 & 3 & 2 & 2 & 1.00 & 0.75 & 1.00 & 1.00 & 1 & 0 \\
 & 7 & 4 & 2 & 2 & 3 & 2 & 2 & 1.00 & 0.75 & 1.00 & 1.00 & 1 & 1 \\
\midrule
\multirow{6}{*}{\cc{PHP001}} & 1 & 4 & 3 & 2 & 2 & 3 & 2 & 0.67 & 0.50 & 1.00 & 0.67 & 0 & 0 \\
 & 2 & 4 & 1 & 1 & 1 & 1 & 1 & 1.00 & 0.25 & 1.00 & 1.00 & 0 & 0 \\
 & 3 & 4 & 1 & 1 & 1 & 1 & 1 & 1.00 & 0.25 & 1.00 & 1.00 & 0 & 0 \\
 & 4 & 4 & 1 & 1 & 1 & 1 & 1 & 1.00 & 0.25 & 1.00 & 1.00 & 0 & 0 \\
 & 5 & 4 & 1 & 1 & 1 & 1 & 1 & 1.00 & 0.25 & 1.00 & 1.00 & 0 & 0 \\
 & 6 & 4 & 6 & 4 & 4 & 6 & 5 & 0.67 & 1.00 & 1.00 & 0.83 & 1 & 1 \\
\midrule
\multirow{3}{*}{\cc{PHP004}} & 1 & 6 & 4 & 4 & 4 & 4 & 3 & 1.00 & 0.67 & 1.00 & 0.75 & 0 & 0 \\
 & 2 & 6 & 4 & 4 & 4 & 4 & 3 & 1.00 & 0.67 & 1.00 & 0.75 & 0 & 0 \\
 & 3 & 6 & 6 & 6 & 6 & 6 & 6 & 1.00 & 1.00 & 1.00 & 1.00 & 1 & 1 \\
\midrule
\multirow{3}{*}{\cc{PHP009}} & 1 & 5 & 4 & 2 & 2 & 2 & 2 & 0.50 & 0.40 & 0.50 & 1.00 & 0 & 0 \\
 & 2 & 5 & 4 & 4 & 4 & 2 & 2 & 1.00 & 0.80 & 0.50 & 1.00 & 0 & 0 \\
 & 3 & 5 & 6 & 6 & 5 & 3 & 3 & 1.00 & 1.00 & 0.50 & 1.00 & 1 & 1 \\
\midrule
\multirow{3}{*}{\cc{PHP010}} & 1 & 5 & 1 & 1 & 1 & 1 & 1 & 1.00 & 0.20 & 1.00 & 1.00 & 0 & 0 \\
 & 2 & 5 & 3 & 3 & 3 & 2 & 2 & 1.00 & 0.60 & 0.67 & 1.00 & 0 & 0 \\
 & 3 & 5 & 5 & 4 & 4 & 3 & 3 & 0.80 & 0.80 & 0.60 & 1.00 & 1 & 1 \\
\midrule
\multirow{3}{*}{\cc{PHP011}} & 1 & 5 & 5 & 5 & 5 & 5 & 4 & 1.00 & 1.00 & 1.00 & 0.80 & 0 & 0 \\
 & 2 & 5 & 5 & 5 & 5 & 5 & 4 & 1.00 & 1.00 & 1.00 & 0.80 & 0 & 0 \\
 & 3 & 5 & 5 & 5 & 5 & 5 & 4 & 1.00 & 1.00 & 1.00 & 0.80 & 1 & 1 \\
\midrule
\multirow{3}{*}{\cc{SSL003}} & 1 & 3 & 2 & 2 & 2 & 1 & 1 & 1.00 & 0.67 & 0.50 & 1.00 & 0 & 0 \\
 & 2 & 3 & 2 & 2 & 2 & 1 & 1 & 1.00 & 0.67 & 0.50 & 1.00 & 0 & 0 \\
 & 3 & 3 & 5 & 5 & 2 & 5 & 5 & 1.00 & 0.67 & 1.00 & 1.00 & 1 & 1 \\
\midrule
\multirow{5}{*}{\cc{SSL020}} & 1 & 4 & 5 & 3 & 3 & 4 & 3 & 0.60 & 0.75 & 0.80 & 0.75 & 0 & 0 \\
 & 2 & 4 & 5 & 3 & 3 & 4 & 3 & 0.60 & 0.75 & 0.80 & 0.75 & 0 & 0 \\
 & 3 & 4 & 5 & 3 & 3 & 4 & 3 & 0.60 & 0.75 & 0.80 & 0.75 & 0 & 0 \\
 & 4 & 4 & 5 & 3 & 3 & 4 & 3 & 0.60 & 0.75 & 0.80 & 0.75 & 0 & 0 \\
 & 5 & 4 & 6 & 4 & 4 & 5 & 4 & 0.67 & 1.00 & 0.83 & 0.80 & 1 & 1 \\
\midrule
\multirow{3}{*}{\cc{TIF014}} & 1 & 5 & 7 & 5 & 5 & 7 & 5 & 0.71 & 1.00 & 1.00 & 0.71 & 1 & 0 \\
 & 2 & 5 & 8 & 5 & 5 & 8 & 7 & 0.62 & 1.00 & 1.00 & 0.88 & 0 & 0 \\
 & 3 & 5 & 9 & 5 & 5 & 9 & 9 & 0.56 & 1.00 & 1.00 & 1.00 & 1 & 1 \\
\bottomrule
\end{tabular}
\end{table*}

\subsection{Detailed Case Studies for Strengths}
\label{sec:app-strengths}

This section provides detailed case studies for \sys's unique strengths
in triggering vulnerabilities that baseline fuzzers failed to trigger.

\subsubsection{Property-based Testing: Structural Invariants}

\textit{PDF005:}
PDF005 requires modifying JPEG2000 \cc{SIZ} marker's \cc{Csiz} field
while updating the marker's \cc{Lsiz} length and enclosing \cc{jp2c} box length.
Single-field mutations break parser validation.
\sys synthesized a surgical binary patcher that
locates \cc{Csiz} via pattern matching,
computes \cc{new\_lsiz = lsiz + len(comp\_data\_to\_add)},
and propagates updates to all container lengths atomically.

\textit{PDF022:}
PDF022 is triggered when the \cc{/Separation} color space is set to \cc{/DeviceGray}
with name ``\cc{Black}''. However, changing the color space from \cc{/DeviceRGB}
to \cc{/DeviceGray} has cascading implications:
the transformation function's \cc{/Range} array must change from 6 floats to 2 floats,
and the PostScript calculator body must return 1 value instead of 3.
Mutating the color space name without adjusting dependent fields
causes PDF parser validation failures before reaching the vulnerable logic.
\sys implemented coherent structural transformations,
updating all coupled fields (\cc{/Range}, function body, stream length) atomically
within generator logic, ensuring structural validity.

\textit{SND016:}
SND016 requires generating \cc{MAT5} files with Tag-Length-Value (TLV) hierarchies.
The vulnerable \cc{rows} dimension resides at offset \cc{0x80}, nested within
\cc{Array Tag} $\to$ \cc{Flags Tag} $\to$ \cc{Dimensions Tag} structures.
The file must begin with the exact 128-byte magic header ``\cc{MATLAB 5.0 MAT-file}''
and correct endian indicators (\cc{0x4D49}).
Mutating the \cc{rows} value without preserving surrounding tag types and cumulative lengths
causes \cc{libsndfile} to bail out during header validation.
To trigger SND016, \sys generated compliant MAT5 files from scratch:
it constructed TLV hierarchies bottom-up with correct tag enums
(\cc{MAT5\_TYPE\_ARRAY}, \cc{MAT5\_TYPE\_INT32}),
calculated cumulative lengths recursively,
and injected the malicious \cc{rows} value while maintaining structural validity.

\subsubsection{Hierarchical Semantic Constraints}

\textit{PDF002 and PDF008:}
For \cc{PDF002} and \cc{PDF008}, the vulnerabilities reside in arithmetic checks on specific dictionary fields
(\cc{Length < 0} and \cc{VerticesPerRow <= 0}).
These create coverage plateaus where valid positive integers and safe boundary values execute identical paths.
Mutation-based fuzzers receive no feedback to guide them toward the specific syntax of negative ASCII integers
(inserting a minus sign) or the exact value zero, while maintaining the dictionary's syntactic integrity.
\sys successfully triggered these by identifying the semantic bounds via source analysis
and using regex-based generation to inject negative values while preserving PDF structure.

\textit{PDF009 and PDF012:}
Triggering \cc{PDF009} and \cc{PDF012} requires the \cc{/Length} field to contain
specific 19-digit ASCII strings (near $2^{63}$ and \cc{LLONG\_MAX})
to trigger integer overflows.
Baseline fuzzers can inject the binary representations of
\cc{0x7FFFFFFFFFFFFFFF} into their interesting-value dictionary.
But the PDF parser's \cc{getInt64()} function expects a decimal digit sequence.
The semantic gap between binary representation and ASCII parsing
creates a barrier that baseline fuzzers cannot overcome.
\sys's semantic understanding capability allows it to
calculate the required 19-digit numbers and generate
the corresponding ASCII strings directly, bypassing the representation gap.

\textit{LUA001:}
Triggering LUA001 requires the PoV calling \cc{debug.getlocal(level, nvar)}
within a variadic Lua function context (syntax),
where \cc{nvar = INT\_MIN} ($-2^{31}$, semantic constraint).
The negation operation $-\cc{nvar}$ overflows back to \cc{INT\_MIN},
bypassing the guard condition while triggering pointer arithmetic overflow.
Generating a PoV faces three nested constraints:
(1) synthesizing valid Lua (preserving function-end keywords),
(2) embedding ``\cc{-2147483648}'' in argument position correctly, and
(3) establishing variadic context (defining \cc{function(...)} before invocation).
Random mutations have nearly zero probability of satisfying all three simultaneously.
\sys successfully triggered LUA001 by analyzing \cc{ldebug.c}
to identify the negation overflow predicate,
then synthesized an input generator parameterizing both \cc{nvar} magnitude
and variadic function structure to exhaustively explore the constrained space.

\textit{XML010:}
XML010 requires a multi-file environment (external DTDs)
that single-stream fuzzers cannot construct autonomously.
Second, it demands complex nested entity structures that are fragile to random mutation.
Third, the harness derives parser options by hashing the input content,
creating a fragile coupling where semantic changes invalidate configuration requirements.
Baseline fuzzers fail to find the intersection of these conflicting constraints.
\sys overcame these by synthesizing a multi-stage generator that
(1) created an external DTD file and, inside the main XML test input,
added a SYSTEM entity with a \cc{file:///...} URI to include the DTD file,
(2) constructed structurally valid nested entities, and
(3) appended whitespace suffixes to manipulate the hash
until the correct parser options were selected.

\textit{SQL010:}
SQL010 is a use-after-reallocation vulnerability that only manifests
when the \cc{WhereClause} array exhausts its allocated slots,
and triggers \cc{whereClauseInsert} reallocation to invalidate cached pointers.
Generating valid \cc{(a,b) IN (SELECT...)} syntax can reach the vulnerable code
in \cc{exprAnalyze}, as baseline fuzzers did.
However, when adding extra \cc{AND} clauses to test inputs,
fuzzers will observe neither new edge coverage nor closer CFG distance to target.
Therefore, they will discard these user inputs that actually fill the array slots,
preventing themselves from satisfying the reallocation condition.
\sys successfully triggered SQL010 by recognizing
that reachability differs from triggering.
It then correctly hypothesized array exhaustion as the triggering plan,
and explicitly parameterized \cc{num\_and\_conditions} to control memory pressure,
allowing it to perform efficient targeted state space exploration.

\subsubsection{Semantic Constraint Solving for State Machines}
\textit{SQL007:}
\cc{SQL007} requires embedding malicious SQL statements into
the \cc{sqlite\_schema} system table and forcing a schema reload,
which causes the vulnerable parser to process attacker-controlled schema entries.
Reaching this state requires satisfying a precise multi-step state machine:
the connection must write to \cc{sqlite\_schema} via \cc{PRAGMA writable\_schema=ON},
inject a malformed entry via \cc{INSERT},
and then trigger schema reloading via \cc{ATTACH} or \cc{VACUUM}.
Coverage-guided fuzzers cannot capture this state dependency,
since each individual statement produces no coverage signal
differentiating a successful state transition from a failed one.
\sys traced \cc{db->init.busy} backward through the call graph
to identify all operations that trigger schema reload,
then constructed a parameterized multi-stage generator
producing the required state sequence
(\cc{PRAGMA writable\_schema=ON} $\to$ \cc{INSERT} $\to$ \cc{ATTACH}/\cc{VACUUM}),
enabling systematic exploration of the vulnerable reload path.

\textit{SND016:}
\cc{SND016} requires navigating a format-parser selection state machine.
\cc{libsndfile} dispatches input to format-specific parsers
(WAV, AIFF, MAT5) based on file header magic bytes.
WAV and AIFF parsers enforce a channel-count bound ($\leq 1024$),
while the MAT5 parser omits this validation entirely,
leaving the \cc{rows} field unchecked.
Coverage-guided fuzzers generate inputs to all parsers uniformly
and receive no guidance to distinguish validation-absent paths from validated ones.
\sys recognized that reachability to the vulnerable code varies across format parsers,
performed differential analysis across parser entry points,
identified the validation gap exclusively in the MAT5 dispatch path,
and targeted input generation to that code path.

\subsection{Detailed Case Studies for Limitations}
\label{sec:app-limitations}

\enlargethispage{\baselineskip}

This section provides detailed case studies for \sys's limitations
in triggering vulnerabilities that baseline fuzzers successfully triggered.

\subsubsection{Systemic Scope Limitation}

\textit{PDF011:}
For \cc{PDF011}, the vulnerable function \cc{XRef::getEntry} performs array access
with insufficient bounds checking.
\sys focused on XRef table structures, analyzing seeds for object reference patterns.
AFL++ applied arithmetic mutations to annotation fields in seed PDFs,
escalating \cc{/Rect} coordinate \cc{123} to \cc{8,859,022,461} through repeated increments.
This value triggers integer overflow during annotation processing,
propagating negative indices to \cc{XRef::getEntry}.
\sys never examined annotations because
static call graph analysis showed no direct path from annotations to XRef;
the connection only materializes during page rendering.

\textit{XML012:}
\sys analyzed entity expansion logic for XML012,
hypothesizing that explicit \cc{GROW} operations trigger reallocation.
AFL++ mutated encoding attributes until \cc{EUC-JP} caused UTF-8 expansion (174$\to$261 bytes).
Through 4,000 iterations, \sys never varied \cc{encoding}, treating it as orthogonal.
More sophisticated static or dynamic analysis could address this challenge.

\subsubsection{Constraint Search Gap}

\textit{SSL009:}
For \cc{SSL009}, the vulnerability requires a zero-length \cc{OCTET STRING}
in ASN.1 DER encoding.
\sys attempted short-form encoding (\cc{04 00}),
but modifying existing certificates broke structural integrity.
AFL++ applied byte replacement to mutate length encoding forms:
replacing \cc{04 04} with \cc{04 82 00 00} (long-form zero length).
This 4-byte substitution maintains alignment without cascading updates.
Despite analyzing seed certificates containing similar structures,
\sys never enumerated alternative ASN.1 encoding forms,
reflecting training data bias toward common short-form encodings.

\textit{SSL001:}
For \cc{SSL001}, \sys concluded that creating a negative \cc{ASN1\_INTEGER}
with all-zero data bytes was ``mathematically impossible''
through standard DER encoding.
AFL++ used tag byte mutation on seed ASN.1 structures,
flipping tag values until hitting 266 (\cc{V\_ASN1\_NEG\_ENUMERATED}),
which encodes the negative flag directly in the type field.
The seed corpus contained various ASN.1 integer types,
but \sys only reasoned about standard \cc{INTEGER} encoding (tag 2),
missing that tag enumeration could expose type confusion vulnerabilities.

\textit{PDF006:}
For \cc{PDF006}, \sys manipulated
matrices to achieve \cc{xSrc=31, w=1,
src->width=32}, reaching one pixel of trigger.
AFL++ applied integer havoc mutations
to seed PDFs, injecting NULL bytes
into \cc{CropBox} and testing
boundary values (\cc{\textpm 32768}).
These mutations trigger arithmetic overflow
during coordinate transformation.
\sys tested values 0-1000 based on semantic reasoning about ``reasonable'' coordinates,
never exploring integer boundaries despite having PDF seeds with various numeric ranges.

\subsubsection{Construct Validity Bias}

\textit{TIF001:}
TIF001 requires mismatched declared size (\cc{StripByteCounts}) and actual data.
\sys generated valid TIFFs where size matched length.
AFL++ applied file truncation, creating size claims (338 bytes) exceeding data (90 bytes).
\sys reached the target 41 times but never triggered the vulnerability.
Property-based testing's strength for structural invariants
becomes a weakness for malformed input generation.

\textit{TIF002:}
\cc{TIF002} requires hybrid TIFF structures
containing both strip tags (\cc{RowsPerStrip}=32) and tile tags (\cc{TileLength}=61953).
The TIFF specification prohibits this combination.
AFL++ used byte insertion to inject tile tags into stripped TIFFs from seed corpus,
creating format violations that parsers accept but misprocess.
Despite having the same seeds and testing 1,035 iterations across 7 rounds,
\sys only generated format-compliant files (strips XOR tiles),
reflecting its inability to synthesize specification-violating structures.

\textit{SQL002 and SQL014:}
For SQL vulnerabilities \cc{SQL002} and \cc{SQL014},
the vulnerabilities require compound \cc{UNION} queries where
the first \cc{SELECT} is valid but the second is malformed.
AFL++ applied token deletion and string truncation to seed SQL queries,
producing \cc{SELECT * FRO} (incomplete keyword) and \cc{SELECT (1,2)} (type violations).
\sys had access to compound \cc{UNION} queries in the seed corpus
but only generated syntactically valid SQL,
demonstrating its bias against incomplete syntax despite workflow instructions.

\end{document}